\documentclass[preprint,showpacs,aps,pre,amsmath,amssymb]{revtex4}

\usepackage{amsmath}
\usepackage{graphicx}
\usepackage{dcolumn}
\usepackage{bm}

\begin{document}
\title{\textbf{Linear Response and Hydrodynamics for Granular Fluids}}
\author{James W. Dufty}
\affiliation{Department of Physics, University of Florida,
Gainesville, FL 32611}
\author{Aparna Baskaran}
\altaffiliation[Present address: ]{Physics Department, Syracuse
University, Syracuse, NY 13244} \affiliation{Department of Physics,
University of Florida, Gainesville, FL 32611}
\author{J. Javier Brey}
\affiliation{F\'{\i}sica Te\'{o}rica, Universidad de Sevilla,
Apartado de Correos 1065, E-41080, Sevilla, Spain}

\date{\today }

\begin{abstract}
A formal derivation of linear hydrodynamics for a granular fluid is
given. The linear response to small spatial perturbations of the
homogeneous reference state is studied in detail using methods of
non-equilibrium statistical mechanics. A transport matrix for
macroscopic excitations in the fluid is defined in terms of the
response functions. An expansion in the wavevector to second order
allows identification of all phenomenological susceptibilities and
transport coefficients through Navier-Stokes order, in terms of
appropriate time correlation functions. The transport coefficients
in this representation are the generalization to granular fluids of
the familiar Helfand and Green-Kubo relations for normal fluids. The
analysis applies to a wide range of collision rules. Important
differences in both the analysis and results from those for normal
fluids are identified and discussed. A scaling limit is described
corresponding to the conditions under which idealized inelastic hard
sphere models can apply. Further details and interpretation are
provided in the paper following this one, by specialization to the
case of smooth, inelastic hard spheres with constant coefficient of
restitution.
\end{abstract}


\maketitle

\affiliation{Department of Physics, University of Florida,
Gainesville, FL 32611}

\affiliation{F\'{\i}sica Te\'{o}rica, Universidad de Sevilla,
Apartado de Correos 1065, E-41080, Sevilla, Spain}

\section{Introduction}

Forty years ago, significant advances in the theory and simulation
of simple atomic fluids were stimulated by the application of exact
methods from non-equilibrium statistical mechanics, namely linear
response and the ``time correlation function method'' \cite{old}.
The results differed from earlier studies based on approximate
kinetic theories in that they are formally exact and closely related
to properties measured in experiments. Subsequently, a great deal
has been learned through the study of appropriate time correlation
functions by theory, simulation, and experiment \cite {Boon80}. In
many respects, the more recent study of granular fluids is poised to
exploit this body of work on normal fluids. Significant advances
have been made in the past decade through the application of
molecular dynamics simulation and kinetic theory. However, although
the generalization of the formal structure for non-equilibrium
statistical mechanics has been described \cite {Brey97,vanN01},
relatively few applications outside of simulation and kinetic theory
\cite{Poschel1,Poschel2,Goldhirsch03} have been given. In
particular, formally exact relations between properties of interest
and appropriate time correlation functions appear restricted to the
simplest cases of tagged particle motion
\cite{Brilliantov00,Dufty00,Dufty01,Dufty02} and liquid structure
\cite{Lutsko01}.

The objective here is to provide a general application of these
formal methods to granular fluids. This is done by studying the
response to small perturbations of a reference homogeneous state,
and using this to extract formally exact expressions for the
hydrodynamic transport coefficients up through Navier-Stokes order.
This is the analogue of the study of excitations about the
equilibrium Gibbs state for normal fluids. In many experimental
conditions of interest, for both normal and granular fluids, the
system is not close to a global homogeneous state. Nevertheless, it
is expected that the reference states studied here are relevant
\emph{locally} for more complex and realistic physical conditions
\cite{Dufty03}. For example, the transport coefficients such as
viscosity and thermal conductivity obtained from linear response,
are the same functions of density and temperature as those in the
associated non-linear equations applicable under more general
conditions. More specifically, the transport coefficients discussed
here are the same as those studied to date for granular fluids using
kinetic theory, but without the limitations or assumptions of those
theories.

The problem of linear response can be formulated at both the level
of phenomenological hydrodynamics and statistical mechanics. In both
cases, the response of the hydrodynamic fields  $y_{\alpha }\left(
\bm{r},t\right) $ (local number density, granular temperature, and
flow velocity), defined as functions of position ${\bm r}$ and time
$t$, to small initial spatial deviations  $\delta y_{\alpha }({\bm
r},0)$ from their values in a homogeneous reference state, is
written in terms of a matrix of response functions $C_{\alpha \beta
}\left( \mathbf{r};t\right) $,
\begin{equation}
\delta y_{\alpha }\left( \mathbf{r},t\right) =\sum_{\beta} \int d
{\bm r}^{\prime}\, C_{\alpha \beta }\left( {\bm r}-{\bm r}^{\prime
};t\right) \delta y _{\beta }\left( {\bm r}^{\prime },0\right) .
\label{1.1}
\end{equation}
These response functions can be calculated approximately using the
linearized phenomenological hydrodynamic equations, to obtain an
explicit result as a function of the parameters (e.g., cooling rate,
pressure, transport coefficients) in those equations. On the other
hand, $C_{\alpha \beta }\left( {\bm r};t\right) $ can be given a
formally exact representation from non-equilibrium statistical
mechanics in terms of time correlation functions evaluated in the
reference homogeneous state. If the hydrodynamic equations are valid
on some length and time scale, then the exact and phenomenological
representations must be the same in that context
\cite{Onsager,Kadanoff,Helfand,McL}. This provides a means for
identifying the phenomenological parameters such as transport
coefficients in terms of exact properties of the correlation
functions, that is a precise link between the macroscopic properties
of interest under experimental conditions and the underlying
fundamental microscopic laws.

There are two parts to this prescription. The first is a
demonstration that the statistical mechanical representation admits
a limit with the same form as that from hydrodynamics, allowing
identification of the hydrodynamic parameters. The second is a proof
that the hydrodynamics dominates all other possible excitations in
this limit. The first part constitutes the usual derivation of
Helfand \cite{Helfand} and Green-Kubo \cite{McL} time correlation
function expressions for transport coefficients in a normal fluid,
and the presentation here is essentially its extension to granular
fluids. The second part is more difficult and remains incomplete
even for normal fluids. The argument there is that the hydrodynamic
fields have a dynamics that persists on the longest time scale,
since they are the densities of conserved quantities and therefore
have infinite relaxation times at infinite wavelengths. Hence it
should be possible to wait for all other excitations to decay,
leaving a space and time scale on which only hydrodynamics remains.
This is the limit in which the Green-Kubo expressions can be
identified and hydrodynamics dominates. A similar assumption is made
here for granular fluids, as discussed further below.

The expressions obtained in this way for the phenomenological
hydrodynamic parameters are still formal, in the sense that the time
correlation function expressions are difficult to evaluate. However,
they depend only on the low frequency, long wavelength limit of the
response functions and, therefore, do not entail the full complexity
of evaluating the complete functions $C_{\alpha \beta }\left( {\bm
r};t\right) $. This approach gives directly exact expressions for
the parameters, without the intermediate practical assumptions and
calculations needed to determine $C_{\alpha \beta }\left( {\bm r}
;t\right) $ first. For example, it is entirely possible that the
usual forms of kinetic theory, and the consequent derivation of
hydrodynamics from them, might fail for granular fluids under some
conditions. Still, hydrodynamics might apply independent of the
kinetic theory basis and the correct expressions for the transport
coefficients would be those obtained from linear response. The
utility of such formal results for both simulation and theory, has
been illustrated recently for granular fluids in the cases of
mobility \cite{Dufty01} and impurity diffusion \cite{Dufty02}.

Although the above prescription for application of linear response
to the hydrodynamics of a granular fluid is simple to state in
general, its implementation in detail requires addressing a number
of differences from the case of normal fluids. In granular fluids,
there is no ''approach to equilibrium'' in the usual sense because
there is no equilibrium Gibbs state, due to the continual loss of
energy by inelastic collisions. Thus Eq.\ (\ref{1.1}) constitutes a
reformulation of Onsager's observation that linear non-equilibrium
regression laws can be studied via equilibrium fluctuations
\cite{Onsager,dGyM69}. It appears that there is a ``universal''
homogeneous state for an isolated granular fluid replacing the Gibbs
state, which is ``normal'' in the sense that all time dependence
occurs through the average energy (or, equivalently, the granular
temperature). This is usually referred to as the Homogeneous Cooling
State (HCS), and it is the reference state about which spatial
perturbations are studied. However, the HCS distribution is not
simply a function of the global invariants nor is it stationary. The
time dependence of this reference state, although occurring only
through the temperature, provides a technical complication in the
application of standard methods of non-equilibrium statistical
mechanics.

Another related complication is the existence of homogeneous state
dynamics (homogeneous perturbations of the HCS) for a granular
fluid, even at the hydrodynamic level. In a normal fluid, the
homogeneous hydrodynamic state is time independent, corresponding to
the global conserved number, energy, and momentum. These dynamical
invariants are the hydrodynamic modes in the extreme long wavelength
limit. The only dynamical response to initial homogeneous
perturbations of the equilibrium state corresponds to
``microscopic'' excitations that decay quickly. Such transient
effects can be eliminated from the response function in Eq.\
(\ref{1.1}) by a suitable choice for the initial perturbation, such
that only the invariants occur in the long wavelength limit. For
normal fluids, this is accomplished in all previous derivations by
choosing an initial \textit{local equilibrium }ensemble. This choice
is both practical (eliminating initial homogeneous transients) and
physically meaningful. The latter refers to the expectation that
each cell in a fluid rapidly approaches the equilibrium state for
that cell, but at its own local density $n({\bm r},t)$, temperature
$T({\bm r},t)$, and flow velocity ${\bm U}({\bm r},t)$. Thus most
initial preparations will quickly approach the local equilibrium
state, and its choice as an initial condition avoids the details of
these short time transients. A similar expectation seems reasonable
for granular fluids, where there should be a rapid approach to a
local HCS. Initial preparation in this state again avoids the
problem of initial transients, but there is still a residual
homogeneous hydrodynamics associated with a uniform perturbation of
all cells. Such excitations correspond to the invariants of a normal
fluid, and again represent the hydrodynamic modes in the long
wavelength limit. They are discussed in the next section at the
level of phenomenological hydrodynamics. In Sec. \ref{sec3}, these
modes are identified in an exact solution to the microscopic
Liouville equation, showing the existence of hydrodynamic modes in
the long wavelength limit and providing the basis for an appropriate
analysis at finite wavelengths.

With the homogeneous hydrodynamics characterized, the residual
dynamics of the response functions determines the parameters
representing effects of spatial gradients. For linear response, it
is sufficient to consider a single Fourier mode with wavevector of
magnitude $k=2\pi /\lambda $, where $\lambda $ is the wavelength of
the spatial perturbation. A formal generator for the dynamics of the
response function is defined and expanded for long wavelengths to
second order in $k$. If the coefficients in this expansion have a
finite limit for long times, they define the corresponding transport
matrix for the phenomenological equations. Accordingly, the
parameters (cooling rate, pressure, and seven transport
coefficients) are given definitions in terms of correlation
functions for the reference HCS in this limit. The analysis is
straightforward but complex, so many of the details are deferred to
appendices. It is perhaps helpful to see an overview of the results
here before getting too immersed in that analysis and associated
notation.

First, the reference HCS is discussed and defined in a
representation where it is a stationary solution to a modified
Liouville equation, analogous to the equilibrium Gibbs stationary
solution. The explicit construction of this solution is a difficult
many-body problem and is not discussed here beyond noting the many
studies of this state by molecular dynamics simulation \cite
{HCSMD}. An exact solution to this Liouville equation for a special
homogeneous perturbation of the HCS is constructed and shown to have
the same dynamics as that from hydrodynamics in the long wavelength
limit. This allows identification of the microscopic hydrodynamic
modes and ``invariants'' for a granular fluid. A formal solution to
the Liouville equation for corresponding perturbations at finite
wavevector is constructed, and the response functions defined. The
expansion in $k$ of the above solution is performed, with
coefficients identified in terms of time correlation functions for
the HCS. The phase functions defining them are fluxes of the usual
conserved densities, and conjugate fluxes associated with densities
of the invariants. In addition, there are correlation functions for
the energy loss function due to the inelastic collisions.

The cooling rate and the pressure in the linear hydrodynamic
equations are identified as specific averages over the HCS
solution. In particular, the pressure is the same average of the
trace of the microscopic stress tensor as for an equilibrium
fluid, but with the equilibrium Gibbs distribution replaced by the
HCS. This determines the dependence of the pressure on density and
temperature. The transport coefficients are of two types, those
associated with the heat and momentum fluxes, and those associated
with the cooling rate. In each case they can be displayed
collectively in matrix form. For the fluxes, they have a
Green-Kubo representation in terms of the above-mentioned fluxes,
\begin{equation}
\Lambda _{\alpha \beta }\left( n,T\right) =\Lambda _{\alpha \beta
}^{(0)}\left( n,T\right) - \lim \int_{0}^{t}dt^{\prime }G_{\alpha
\beta }(n,T;t^{\prime }).  \label{1.2}
\end{equation}
The first term on the right hand side is a time independent
correlation function for the HCS. It vanishes for normal fluids with
continuous potentials of interaction, and occurs here due to the
inelasticity of the collisions. Such a term can occur even in the
elastic limit for singular forces, such as hard spheres
\cite{Dufty04}. The limit indicated in the second term is the usual
thermodynamic limit of large volume $V$ and particle number $N$,
followed by the limit of large time. The integrand $G_{\alpha \beta
}(n,T;t)$ is a flux-flux correlation function
\begin{equation}
G_{\alpha \beta }(n,T;t)\equiv \frac{1}{V}\sum_{\lambda}\int d\Gamma
{\Phi}_{\alpha }\left( \Gamma ;n,T\right) \mathcal{U}_{\beta \lambda
}\left( t,T\right) \Upsilon _{\lambda }\left( \Gamma ;n,T\right) .
\label{1.3}
\end{equation}
The first flux $\Phi_{\alpha }$, is one of those associated with the
usual densities of number, energy, and momentum. The second flux $
\Upsilon _{\lambda }$, is one of those associated with the densities
for the new invariants. Both kind of fluxes are functionals of the
phase point $\Gamma$. The evolution operator for the dynamics
$\mathcal{U}_{\alpha \beta }\left( t,T\right) $ is the usual
Liouville dynamics, but now compensated for the homogeneous dynamics
of collisional cooling and homogeneous perturbation. It has an
invariant subspace $\{\Psi_{\alpha}(\Gamma;n,T)\}$,
\begin{equation}
\sum_{\beta} \mathcal{U}_{\alpha \beta }\left( t,T\right) \Psi
_{\beta }\left( \Gamma ;n,T\right) =\Psi _{\alpha }\left( \Gamma
;n,T\right). \label{1.4}
\end{equation}
Such a contribution would invalidate the limit in Eq.\ (\ref{1.2}),
but it is demonstrated that the fluxes $\Upsilon _{\lambda }$ are
orthogonal to this subspace. This corresponds to the ``subtracted
fluxes'' in the Green-Kubo expressions for normal fluids \cite{McL},
which are recovered from these results in the elastic limit. The
transport coefficients associated with the cooling rate have a
similar form, with $\Phi_{\alpha }$ replaced by a phase function
representing the energy loss from the inelastic collisions.

In both cases, the time correlation functions can be expressed
exactly as a time derivative. Performing the time integral in Eq.\
(\ref{1.2}) leads to an alternative representation. In fact, three
representations for the transport coefficients are obtained here.
They correspond to the familiar results for the diffusion
coefficient of an impurity in a normal fluid,
\begin{equation}
D=- \lim \frac{1}{d}\, \frac{\partial}{\partial t}\left\langle {\bm
q}_{0}\cdot {\bm q} _{0}(t)\right\rangle_{c} = - \lim
\frac{1}{d}\left\langle {\bm q}_{0}\cdot {\bm
v}_{0}(t)\right\rangle_{c} =\lim \frac{1}{d}\int_{0}^{t}dt^{\prime
}\left\langle {\bm v}_{0}\cdot {\bm v}_{0}(t^{\prime
})\right\rangle_{c}, \label{1.5}
\end{equation}
where ${\bm q}_{0}$ and ${\bm v}_{0}$ are the position and velocity
of the impurity, $d$ is the dimension of the system, and the angular
brackets denote equilibrium averages. The first is the Einstein
relationship, indicating that the mean square displacement grows
linearly with time for long times. It was subsequently generalized
to other transport coefficients by Helfand \cite{Helfand}, and it is
usually referred to as  the Einstein-Helfand representation. The
other two representations are the intermediate Helfand, and
Green-Kubo forms, respectively.

The derivation of these formal results is accomplished with few
restrictions on the dynamics in phase space: deterministic,
Markovian, and invertible. This allows a wide range of inelastic
collision rules currently used for granular fluids, from inelastic
hard spheres to soft viscoelastic potentials. However, to expose
further details of the expressions for the transport coefficients
obtained here, a companion paper following this presentation is
specialized to the simplest case of smooth, inelastic hard spheres
with constant coefficient of restitution \cite{BDB06,Ba06}. Further
discussion of the formal results is deferred to the final section.

\section{Phenomenological Hydrodynamics}

\label{sec2} The purpose of this section is two-fold. First, the
phenomenological hydrodynamic equations are recalled and the
unknown parameters (pressure, cooling rate, transport
coefficients) indicated. A special solution for spatially
homogeneous states is obtained, and the hydrodynamic equations are
linearized about that state for small spatial perturbations. A
microscopic representation of these equations is the objective of
subsequent sections. The second purpose is to characterize the
dynamics to be expected from solutions to these equations.
Specifically, two features new to granular fluids are an inherent
time dependence of the coefficients due to the cooling of the
reference state, and a non-trivial dynamics associated with
homogeneous perturbations of the homogeneous state. In the
subsequent sections, identifying the component of linear response
associated with only spatial perturbations requires taking
explicit account of these two types of homogeneous dynamics.

Hydrodynamics is a closed description for the dynamics of a fluid in
terms of the number density $n\left( {\bm r},t\right) $, the energy
density $ e\left( {\bm r},t\right) $, and the momentum density ${\bm
g}\left({\bm r},t\right) $. The starting point for identifying such
a description for a granular fluid is the exact macroscopic balance
equations for these densities. However, as for normal fluids, it is
usual to replace the energy density and momentum density by the
temperature $T\left( {\bm r},t\right) $ and flow velocity ${\bm
U}\left( {\bm r},t\right) $ as the hydrodynamic variables, together
with the number density. This is accomplished through the
definitions
\begin{equation}
e\left( {\bm r},t\right) \equiv \frac{1}{2}mn\left( {\bm
r},t\right) U^{2}\left( {\bm r},t\right) +e_{0}\left[ n\left( {\bm
r},t\right) ,T\left( {\bm r},t\right) \right], \label{2.0}
\end{equation}
\begin{equation}
{\bm g}\left( {\bm r},t\right) \equiv mn\left( {\bm r},t\right)
{\bm U}\left( {\bm r},t\right). \label{2.0a}
\end{equation}
Here $m$ is the mass of a particle and  $e_{0}\left( n,T\right) $
is some specified function of $n$ and $T$. The two most common
choices are $e_{0}\left( n,T\right) =dnT/2$, where $d$ is the
dimension of the system, or $ e_{0}\left( n,T\right) =e_{e}\left(
n,T\right)$, the thermodynamic function for the corresponding
equilibrium fluid. The former is common in applications of
computer simulations (note that the Boltzmann constant has been
set equal to unity), while the latter is the historical choice in
most formulations of hydrodynamics. Both definitions coincide for
the special case of hard spheres. For normal and also for granular
fluids, the choice made constitutes a \emph{definition} of
temperature for non-equilibrium states and has no \emph{a priori}
thermodynamic implications. The exact macroscopic balance
equations in terms of $n$, $T$, and ${\bm U}$ are
\begin{equation}
D_{t}n+n{\bm \nabla} \cdot {\bm U}=0,  \label{2.1}
\end{equation}
\begin{equation}
D_{t}U_{i}+(mn)^{-1}\sum_{j} \frac{\partial}{\partial r_{j}}
P_{ij}=0, \label{2.2}
\end{equation}
\begin{equation}
\left( \frac{\partial e_{0}}{\partial T}\right)_{n}\left(
D_{t}+\zeta \right) T+\left[ e_{0}-n \left( \frac{\partial
e_{0}}{\partial n}\right)_{T}\right] \nabla \cdot {\bm U}+\sum_{i}
\sum_{j} P_{ij}\frac{\partial U_{i}}{\partial r_{j}} +\nabla \cdot
\mathbf{q}  =0,  \label{2.3}
\end{equation}
where $D_{t} \equiv \partial /\partial t+\mathbf{U}\cdot \nabla $ is
the material derivative, $\zeta $ is the cooling rate due to the
energy loss from the interaction between granular particles,
$\mathbf{q}$ is the heat flux, and $P_{ij}$ is the pressure tensor.
These equations have the same form as those for a normal fluid,
except for the presence of the term involving the cooling rate
$\zeta $ in the equation for the temperature.

The above balance equations are not a closed set of equations until
${\bm q}$, $P_{ij}$, and $\zeta $ are specified as functionals of
the hydrodynamic fields, i.e., their space and time dependence
occurs entirely through these fields. This happens for normal fluids
on length and time scales long compared to the mean free path and
mean free time, respectively, and similar conditions may be assumed
for granular fluids as well. If, furthermore, the state of the
system is such that the spatial variation of the fields is smooth,
then an expansion of these functionals in gradients of the fields
can be performed. From fluid symmetry, the results to first order in
the gradients must have the form:
\begin{equation}
P_{ij}= p(n,T)\delta _{ij}-\eta (n,T)\left( \frac{\partial U_{i}}{
\partial r_{j}}+\frac{\partial U_{j}}{\partial
r_{i}}-\frac{2}{d}\delta _{ij}{\bm \nabla} \cdot {\bm U} \right)
-\kappa (n,T)\delta _{ij}{\bm \nabla \cdot {\bm U}}+ \ldots ,
\label{2.5}
\end{equation}
\begin{equation}
{\bm q} = -\lambda (n,T){\bm \nabla }T-\mu (n,T) {\bm \nabla }n+
\ldots,  \label{2.6}
\end{equation}
\begin{equation}
\zeta = \zeta _{0}\left( n,T\right) +\zeta ^{U}(n,T){\bm \nabla}
\cdot {\bm U} + \zeta ^{T}(n,T)\nabla ^{2}T+\zeta ^{n}(n,T)\nabla
^{2}n+ \ldots, \label{2.4}
\end{equation}
where $\delta_{ij}$ is the Kronecker delta symbol and the dots
denote terms proportional to higher powers or higher degree spatial
derivatives of the hydrodynamic fields than those written
explicitly. These expressions represent the ``constitutive
equations'' which, together with the macroscopic balance equations,
give Navier-Stokes order hydrodynamics for a granular fluid. Note
that the cooling rate is required to second order in the gradients,
while the pressure tensor and heat flux are required only to first
order. The pressure tensor has the same form as Newton's viscosity
law for a normal fluid, while the expression for the heat flux is a
generalization of Fourier's law \cite{BDKyS98,SMyR99}.

The expressions (\ref{2.5})-(\ref{2.4}) include unspecified
functions: the pressure $p(n,T)$, the zeroth order in the
gradients cooling rate $\zeta _{0}\left( n,T\right) $, the
transport coefficients from the cooling rate $\zeta ^{U}(n,T)$,
$\zeta ^{T}(n,T)$, $\zeta ^{n}(n,T)$, the shear viscosity $\eta
(n,T)$, the bulk viscosity $\kappa (n,T)$, the thermal
conductivity $\lambda (n,T)$, and a new heat flux coefficient $\mu
(n,T)$. All of these must be provided by experiment or by the
theoretical justification of this phenomenology.

Although the Navier-Stokes equations are based on the small
gradient forms for the constitutive equations, it does not mean
that they are limited to systems close to a global homogeneous
state. Thus, they can be applicable locally over domains larger
than the mean free path, but the hydrodynamic fields may still
vary significantly throughout the system. Consequently, a wide
range of experimental and simulation conditions for granular
fluids have been well-described by the Navier-Stokes equations
\cite{BRMC99,Brey,Caldera,Swinney}.

The spatially homogeneous solution to Eqs.\
(\ref{2.1})-(\ref{2.3}) for an isolated system (e.g., periodic
boundary conditions) is
\begin{equation}
n\left( {\bm r},t\right) =n_{h},\quad {\bm U}\left( {\bm r}
,t\right) ={\bm U}_{h},\quad T\left( {\bm r},t\right) =T_{h}(t),
\label{2.7}
\end{equation}
where $T_{h}(t)$ is the solution to
\begin{equation}
\left\{ \frac{\partial}{\partial t}+\zeta _{0}\left[
n_{h},T_{h}\left( t\right) \right] \right\} T_{h}\left( t\right)
=0.  \label{2.7aa}
\end{equation}
For simplicity, it will be often considered that ${\bm U}_{h}=0$ in
the following, something that can always been achieved by means of
the appropriate Galilean transformation. All time dependence of this
state occurs through the homogeneous temperature $T_{h}\left(
t\right) $, so this is referred to as the homogeneous cooling state
(HCS) \cite{Ha83}. Once the functional form for $\zeta _{0}\left[
n_{h},T_{h}\left( t\right) \right] $ has been specified, the first
order nonlinear equation (\ref{2.7aa}) can be solved by direct
integration for a given initial condition.

Now consider small spatial perturbations of the HCS, assuming that
$T_{h}\left( t\right) $ is known,
\begin{equation}
y_{\beta }\left( {\bm r},t\right) =y_{\beta ,h} +\delta y_{\beta
}\left( {\bm r},t\right) ,\quad \left\{ y_{\beta} \right\} \equiv
\left\{ n,T, {\bm U}\right\}, \label{2.7aaa}
\end{equation}
with $\delta y_{\beta }\left( \mathbf{r},t\right) $ sufficiently
small that nonlinear terms in the hydrodynamic equations can be
neglected. The resulting linear equations have coefficients
independent of ${\bm r}$ so the differential equations can be
given an algebraic representation by means of the Fourier
transformation,
\begin{equation}
\delta \widetilde{y}_{\beta }({\bm k},t)=\int d{\bm r}\, e^{i{\bm k}
\cdot {\bm  r}}\delta y_{\beta }({\bm r},t). \label{2.7c}
\end{equation}
Moreover, the components of the flow velocity are separated into a
longitudinal component relative to ${\bm k}$, $\delta
\widetilde{U}_{\parallel}= \hat{\bm k} \cdot \delta \widetilde{\bm
U}$, and $d-1$ transverse components $\delta
\widetilde{U}_{\perp,i}=\hat{\bm e}_{i} \cdot \delta \widetilde{\bm
U}$, where $\hat{\bm k} \equiv {\bm k}/k$ and $ \left\{ \hat{\bm
e}_{i}; i=1, \ldots, d-1 \right\}$ are a set of $d$ pairwise
orthogonal unit vectors. Therefore, from now on it is
\begin{equation}
\left\{ \widetilde{y}_{\beta } \right\} \equiv \left\{
\widetilde{n},\widetilde{T}, \widetilde{U}_{\parallel},
\widetilde{{\bm U}}_{\perp} \right\}, \label{2.7b}
\end{equation}
with $\delta \widetilde{{\bm U}}_{\perp}$ denoting the set of the
$d-1$ components $\delta \widetilde{{U}}_{\perp,i}$. The linearized
hydrodynamic equations then have the form
\begin{equation}
\left\{ \partial _{t}+\mathcal{K}^{hyd}\left[ n_{h},T_{h}(t);{\bm
k} \right] \right\} \delta \widetilde{y}\left( {\bm k},t\right)
=0, \label{2.7a}
\end{equation}
where a $d+2$ dimensional matrix notation has been introduced for
simplicity and $\partial_{t} \equiv \partial / \partial t$. The
transport matrix $\mathcal{K}^{hyd}$ is identified as being block
diagonal, with decoupled longitudinal and transversal submatrices,
\begin{equation}
\mathcal{K}^{hyd}= \left(
\begin{array}{cc}
\mathcal{K}_{1}^{hyd}  & 0 \\
0   & \mathcal{K}_{2}^{hyd}
\end{array}
\right), \label{2.14}
\end{equation}
\begin{equation}
\mathcal{K}^{hyd}_{1}=\left(
\begin{array}{ccc}
0 & 0 & -in_{h}k \\
\frac{\partial \left( \zeta _{0}T_{h}\right) }{\partial
n_{h}}+\left( \frac{\mu }{e_{0,T}} -\zeta ^{n}T_{h}\right) k^{2} &
\frac{\partial \left( \zeta _{0}T_{h}\right) }{\partial
T_{h}}+\left( \frac{\lambda}{e_{0,T}}-\zeta ^{T}T_{h}\right) k^{2}
& -i  \left( \zeta ^{U}T_{h}+\frac{h-e_{0,n}n_{h}}{e_{0,T}} \right)k \\
-\frac{i p_{n}k}{n_{h}m} & -\frac{i p_{T} k}{n_{h}m} &
\frac{1}{n_{h}m}\left[ \frac{2(d-1)}{ d}\eta +\kappa \right] k^{2}
\end{array}
\right) ,  \label{2.14a}
\end{equation}
\begin{equation}
\mathcal{K}^{hyd}_{2}= \frac{\eta}{n_{h}m}\, k^{2} I. \label{2.15}
\end{equation}
In the above expressions, $ I$ is the unit matrix of dimension
$d-1$. Moreover, $h \equiv e_{o}+p$ and the following short
notations have been introduced:
\begin{equation}
e_{0,n} \equiv \left( \frac{\partial e_{0}}{\partial
n}\right)_{T}, \quad e_{0,T} \equiv  \left( \frac{\partial
e_{0}}{\partial T}\right)_{n},\quad p_{n} \equiv
\left(\frac{\partial p}{\partial n} \right)_{T}, \quad p_{T}
\equiv \left(\frac{\partial p}{
\partial T}\right)_{n}.  \label{2.15a}
\end{equation}
It is understood in Eqs. (\ref{2.14a}) and (\ref{2.15}) that all the
quantities are to be evaluated at $n= n_{h}$ and $T= T_{h}\left(
t\right)$.

There are $d+2$ independent solutions of Eqs. (\ref{2.7a}) in terms
of which the general solution to the initial value problem can be
expressed. These are the hydrodynamic modes. For normal fluids, they
are simply the eigenfunctions of the matrix $\mathcal{K}^{hyd}$ and
the hydrodynamic excitations are exponential in time with the
corresponding eigenvalues. Here, the identification is somewhat less
direct due to the dependence of $ \mathcal{K}^{hyd}$ on $T_{h}(t)$.
For example, the shear modes are proportional to
\begin{equation}
\exp\left\{ -\int^{t}dt^{\prime }\, \frac{\eta \left[
n_{h},T_{h}(t^{\prime })\right]}{n_{h}m} k^{2} \right\},
\label{2.16}
\end{equation}
and their time dependence is no longer exponential.

The general solution to the initial value problem can be written in
terms of a matrix of response functions (Green's function matrix)
corresponding to the representation in Eq.\ (\ref{1.1}),
\begin{equation}
\delta \widetilde{y}\left({\bm k},t\right)
=\widetilde{C}^{hyd}\left[ n_{h},T_{h}(t);{\bm k},t\right] \delta
\widetilde{y}\left( {\bm k} ,0\right) ,  \label{2.17}
\end{equation}
which is determined here from the linearized hydrodynamic
equations,
\begin{equation}
\left\{ \partial _{t}+\mathcal{K}^{hyd}\left[ n_{h},T_{h}(t);{\bm k}
\right] \right\} \widetilde{C}^{hyd}\left[ n_{h},T_{h}(t);{\bm
k},t\right]=0, \label{2.18}
\end{equation}
\begin{equation}
\widetilde{C}_{\alpha \beta }^{hyd}\left[ n_{h},T_{h}(0); {\bm
k},0\right] =\delta _{\alpha \beta }.
\end{equation}
Conversely, if $\widetilde{C}^{hyd}\left[ n_{h},T_{h}(t); {\bm
k},t\right]$ were given, the parameters of the transport matrix
$\mathcal{K} ^{hyd}\left[ n_{h},T_{h}(t);{\bm k}\right] $ could be
determined from
\begin{equation}
\mathcal{K}^{hyd}\left[ n_{h},T_{h}(t);{\bm k}\right] =-\left\{
\partial _{t}\widetilde{C}^{hyd}\left[
n_{h},T_{h}(t);{\bm k},t\right] \right\} \widetilde{C}^{hyd
-1}\left[ n_{h},T_{h}(t);{\bm k},t \right] . \label{2.19}
\end{equation}
This exposes the formally exact approach to be developed in the next
sections to identify the parameters of $\mathcal{K}^{hyd}\left[
n_{h},T_{h}(t);{\bm k}\right] $. First, an exact response function
$\widetilde{C}$ is defined in place of $\widetilde{C}^{hyd}$ for
$\delta \widetilde{y}\left( {\bm k},t\right) $. Next, a transport
matrix is defined as in Eq.\ (\ref{2.19}) with $\widetilde{C}
^{hyd}$ replaced by $\widetilde{C}$. On the long time and small $
{\bm k}$ scales for hydrodynamics, this expression must agree with $
\mathcal{K}^{hyd}\left[ n_{h},T_{h}(t);{\bm k}\right] $, providing a
formal definition of the hydrodynamic parameters. Indeed, for normal
fluids this procedure gives the familiar Helfand and Green-Kubo
expressions for the transport coefficients in terms of time
correlation functions in the equilibrium reference state.

There are two immediate complications for direct extension of this
approach to granular fluids. The first is the parametrization of
$\mathcal{K} ^{hyd}\left[ n_{h},T_{h}(t);{\bm k}\right] $ by the
time dependent temperature $T_{h}(t)$. For example, the transport
coefficients are functions of this temperature. Since $T_{h}(t)$ is
determined autonomously from Eq.\ (\ref{2.7aa}), it is useful to
suppress this dynamics by the definitions
\begin{eqnarray}
\mathcal{K}^{hyd}\left[ n_{h},T_{h}(t);{\bm k}\right] &=&\left[
\mathcal{K }^{hyd}\left( n,T;{\bm k}\right)
\right]_{n=n_{h},T=T_{h}(t)},
\notag \\
\widetilde{C}^{hyd}\left[ n_{h},T_{h}(t);{\bm k},t\right]
&=&\left[ \widetilde{C}^{hyd}\left( n,T;{\bm k},t\right)
\right]_{n=n_{h},T=T_{h}(t)}.  \label{2.20}
\end{eqnarray}
Thus $\mathcal{K}^{hyd}\left( n,T;{\bm k}\right) $ and
$\widetilde{C} ^{hyd}\left( n,T;{\bm k},t\right) $ are the
universal function associated with a general class of reference
states. However, in this form $t$ and $T$ become independent
variables so, for example, the equation for $\widetilde{C}
^{hyd}\left( n,T;{\bm k},t\right) $ is not Eq.\ (\ref{2.18}) but
\begin{equation}
\left[ \partial _{t}-\zeta _{0}\left( n,T\right) T
\frac{\partial}{\partial T}+\mathcal{K} ^{hyd}\left( n,T;{\bm
k}\right) \right] \widetilde{C}^{hyd}\left( n,T; {\bm k},t\right)
=0, \label{2.21}
\end{equation}
again with the initial condition  $\widetilde{C}_{\alpha \beta
}^{hyd}\left( n,T; {\bm k},0\right) =\delta _{\alpha \beta }$. The
time derivative is now at constant $T$. The new term with the
temperature derivative in Eq.\ (\ref{2.21}) is the generator for
the dynamics of $T_{h}(t)$ in the sense that for any arbitrary
function $X(T)$ it is
\begin{equation}
X\left[T_{h}(t;T) \right]=\exp \left[-t\zeta _{0}\left(
n_{h},T\right) T \frac{\partial}{\partial T}\right] X(T),
\label{2.22}
\end{equation}
where $T_{h}(t;T)$ is the solution to Eq.\ (\ref{2.7aa}) with
$T_{h}(0)=T$. The proof is given in Appendix \ref{ap1}. Equation
(\ref{2.21}) now has time independent coefficients, but at the
price of introducing a new independent variable $T$. Still, it is
most convenient to work with the generic forms $
\mathcal{K}^{hyd}\left( n,T;{\bm k}\right) $ and $\widetilde{C}
^{hyd}\left( n,T;{\bm k},t\right) $ and the counterpart of Eq.\
(\ref{2.19}),
\begin{equation}
\mathcal{K}^{hyd}\left( n,T;{\bm k}\right) =-\left\{ \left[
\partial_{t}-\zeta _{0}\left( n,T\right) T \frac{\partial}{\partial T} \right] \widetilde{C}
^{hyd}\left( n,T;{\bm k},t\right) \right\} \widetilde{C}^{hyd
-1}\left( n,T; {\bm k},t\right) . \label{2.23}
\end{equation}

The second complication is the existence of a non-zero generator
for homogeneous (zero wave vector) dynamics, i.e.
$\mathcal{K}^{hyd}\left( n,T;{\bm 0}\right) \neq 0$, or $
\widetilde{C}^{hyd}_{\alpha \beta}\left( n,T;{\bm 0},t\right) \neq
\delta_{\alpha \beta}$. The latter can be calculated directly to
give (see Appendix \ref{ap1}):
\begin{equation}
\widetilde{C}^{hyd}\left( n,T;{\bm 0},t\right) =\left(
\begin{array}{cc}
\widetilde{C}^{hyd}_{1} & 0  \\
0 & I
\end{array}
\right), \label{2.23a}
\end{equation}
where
\begin{equation}
\widetilde{C}^{hyd}_{1}\left( n,T;{\bm 0},t\right)= \left(
\begin{array}{ccc}
1 & 0 & 0 \\
\left ( \frac{\partial T}{\partial n}\right)_{T\left( -t;T\right)
} & \left( \frac{\partial T}{\partial T\left( -t;T\right)}\right)_{n} & 0  \\
0 & 0 & 1
\end{array}
\right)   \label{2.23aa}
\end{equation}
and $I$ is again the unit matrix of dimension $d-1$. The
interpretation of the above result is clear when evaluated at $n=
n_{h}$, $T=T_{h}(t)$, where the non trivial matrix elements become
\begin{eqnarray}
\widetilde{C}_{21}^{hyd}\left[ n_{h},T_{h}(t);{\bm 0},t\right] & =
& \left(\frac{
\partial T_{h}(t)}{\partial n_{h}}\right)_{T_{h}(0)}, \nonumber \\
\widetilde{C}_{22}^{hyd}\left[ n_{h},T_{h}(t);{\bm 0},t\right] & =
& \left( \frac{\partial T_{h}(t)}{\partial T_{h}\left( 0\right)
}\right)_{n_{h}}.  \label{2.23b}
\end{eqnarray}
This is just the linear response of the solution to Eq.\
(\ref{2.7aa}) due to changes in the initial conditions. It is the
additional dynamics of the temperature beyond that of $T_{h}(t)$,
due to initial homogeneous density and temperature perturbations, as
the system attempts to approach a new HCS. Note that this it is a
hydrodynamic excitation, and is distinct from the rapid homogeneous
relaxation of other microscopic modes on a much shorter time scale
outside the scope of hydrodynamics. Hence for granular fluids the
hydrodynamic modes cannot be identified simply as those that vanish
for $ k\rightarrow 0,$ as for a normal fluid. Instead, they are
identified as those modes whose homogeneous dynamics becomes that of
Eq.\ (\ref{2.23a}) for $ k\rightarrow 0$. This is done in the next
section.

\section{Statistical Mechanics}

\label{sec3} Consider a volume $V$ enclosing $N$ particles that
interact and they lose energy as a result of this interaction. Also
suppose that the interactions are specified in such a way that if
the positions and velocities of each of the particles are given at a
time $t_{0}$, then there exists a well defined trajectory for the
evolution of the system for all times both earlier and later than
$t_{0}$. The microscopic initial state of the system is entirely
determined by the positions and velocities of all particles, denoted
by a point in a $2dN$ dimensional phase space $\Gamma \equiv \{{\bm
q}_{r},{\bm v}_{r};r=1,\ldots, N\}$. The state of the system at a
later time $t$ is completely characterized by the positions and
velocities of all particles at that time $\Gamma _{t}\equiv \left\{
{\bm q}_{r}(t),{\bm v}_{r}(t);r=1,\ldots, N\right\} $. In summary,
the dynamics is Markovian and invertible.

The statistical mechanics for this system is comprised of the
dynamics just described, a macrostate specified in terms of a
probability density $\rho (\Gamma )$, and a set of observables
(measurables). The expectation value for an observable $A$ at time
$t>0$,  given a state $\rho (\Gamma )$ at $t=0$, is defined by
\begin{equation}
\langle A(t);0\rangle \equiv \int d\Gamma \rho (\Gamma )A(\Gamma
_{t}). \label{3.1}
\end{equation}
The notation $A(\Gamma _{t})$ indicates the function of a given
phase point shifted forward in time along a trajectory.
Equivalently, this can be considered as a function of the initial
phase point and the time, $A(\Gamma ,t)$. This second
representation allows the introduction of a generator $L$ for the
time dependence defined by
\begin{equation}
\langle A(t);0\rangle =\int d\Gamma \,\rho (\Gamma )e^{tL}A(\Gamma
),\label{3.2}
\end{equation}
where the explicit form of $L$ is determined by the specific rule
chosen for the interactions among the particles. This is left
unspecified at the moment. Due to the assumption of invertibility
made above, the dynamics can be transferred from the observable
$A(\Gamma )$ to the state $\rho (\Gamma )$, by defining an adjoint
generator $\overline{L}$,
\begin{equation}
\int d\Gamma \,\rho (\Gamma )e^{tL}A(\Gamma )\equiv \int d\Gamma
\,\left[ e^{-t\overline{L}}\rho (\Gamma )\right] A(\Gamma )\equiv
\int d\Gamma \,\rho (\Gamma ,t)A(\Gamma ).  \label{3.3}
\end{equation}
The form generated by $\overline{L}$ is referred to as Liouville dynamics.
This equivalence \ of observable and state dynamics is expressed in the
above notation as
\begin{equation}
\langle A(t);0\rangle =\langle A;t\rangle. \label{3.3aa}
\end{equation}
In summary, the dynamics of phase functions is governed by an equation of
the form
\begin{equation}
\left( \partial _{t}-L\right) A(\Gamma ,t)=0,  \label{3.5}
\end{equation}
and that of the probability distribution in phase space by a
Liouville equation
\begin{equation}
\left( \partial _{t}+\overline{L}\right) \rho (\Gamma ,t)=0,
\label{3.6}
\end{equation}
where $L$ and $\overline{L}$ are time independent operators.

Time correlation functions are defined in a similar way:
\begin{equation}
C_{AB}(t) \equiv \langle  A(t)  B(0);0\rangle \equiv \int d\Gamma\,
\rho (\Gamma )  A(\Gamma _{t})B(\Gamma ). \label{3.7}
\end{equation}
In terms of the generators introduced above, the correlation
functions can be written
\begin{equation}
C_{AB}(t)=\int d\Gamma\,  \rho (\Gamma )\left[ e^{tL} A(\Gamma
)\right] B(\Gamma ),  \label{3.9}
\end{equation}
or, equivalently,
\begin{equation}
C_{AB}(t)=\int d\Gamma  A(\Gamma )e^{-t\overline{L}}\left[ \rho
(\Gamma )B(\Gamma )\right] .  \label{3.10}
\end{equation}
Further comment on these generators and some examples are given in
Appendix \ref{ap2}.

\subsection{Homogeneous Reference State}

In contrast to normal fluids, there is no stationary solution to the
Liouville equation (\ref{3.6}) for an isolated granular fluid,
because the average energy of the system decreases with time,
\begin{equation}
\partial _{t}\langle E;t\rangle =\langle LE;t\rangle \leq 0,  \label{4.0}
\end{equation}
where $E(\Gamma)$ is the phase function corresponding to the total
energy and the equal sign corresponds to the elastic limit. This
loss of energy due to interactions among the particles is a primary
characteristic of granular fluids. It is convenient to introduce a
granular temperature instead of the average energy in the same way
as is done in (\ref {2.0}) for the phenomenological equations. \ For
a homogeneous state, the flow velocity $\mathbf{U}$ is uniform and
can be chosen to vanish by an appropriate Galilean transformation,
as already mentioned. The temperature definition for the homogeneous
state then is
\begin{equation}
e_{0}\left[ n,T(t)\right] \equiv \frac{1}{V}\langle E;t\rangle ,
\label{4.3}
\end{equation}
and Eq.\ (\ref{4.0}) becomes
\begin{equation}
\partial _{t}T(t)=-\zeta \left( t\right) T(t),  \label{4.3a}
\end{equation}
where the ``cooling rate'' is identified as
\begin{equation}
\zeta \left( t\right) =-\frac{1}{V T} \left( \frac{\partial
T}{\partial e_{0}}\right)_{n} \langle {LE};t\rangle \geq 0.
\label{4.3b}
\end{equation}

Upon writing the last inequality, it has been taken into account
that for both choices of the function $e_{0}(n,T)$  discussed in the
previous section,  $e_{0}$ is an increasing function of $T$. The
above shows that there is no ``approach to equilibrium'' for a
granular fluid, except in the elastic collision limit where $\zeta
\left( t\right) =0$, as no such stationary state exists for an
isolated system. Consequently, there is a large class of dynamical,
homogeneous states depending on the initial preparation. However, as
in a normal fluid, it is expected (and observed in computer
simulations \cite{HCSMD}) that there is a rapid relaxation of
velocities after a few collisions to a ``universal'' state whose
entire time dependence occurs through the cooling temperature. This
special state is called the homogeneous cooling state (HCS), since
it is the microscopic analog of the hydrodynamic HCS described in
Sec. \ref{sec2} above. The probability density for the HCS then
becomes \cite{Brey97,vanN01}
\begin{equation}
\rho _{h}(\Gamma ,t)=\rho _{h}\left[ \Gamma ;n_{h},T_{h}(t) \right],
\label{4.3c}
\end{equation}
where $n_{h}$ is the uniform, time independent density and
$T_{h}(t)$ the decreasing temperature, consistently with the
notation used in Sec.\ \ref{sec2}. Because all time dependence
occurs through $T_{h}(t)$, it follows that $\zeta_{h} \left(
t\right) = \zeta _{0}\left[ n_{h},T_{h}(t)\right] $, whose form is
obtained from
\begin{equation}
\zeta _{0}\left( n,T\right) \equiv -\frac{1}{V T} \left(
\frac{\partial T}{\partial e_{0}}\right)_{n}\int d\Gamma\,  \rho
_{h}(\Gamma ;n,T)LE\left( \Gamma \right) .  \label{4.3d}
\end{equation}
The functional form of $\rho _{h}(\Gamma ;n,T,{\bm U})$ is
determined self-consistently by the Liouville equation (\ref{3.6})
or, equivalently, using the results in Appendix \ref{ap1},
\begin{equation}
\overline{\mathcal{L}}_{T}\rho _{h}(\Gamma ;n,T,{\bm U})=0,
\label{4.4}
\end{equation}
with the definition
\begin{equation}
\overline{\mathcal{L}}_{T}\equiv -\zeta _{0}\left( n,T\right) T
\frac{\partial}{\partial T}+ \overline{L}.  \label{4.4a}
\end{equation}
Here, a homogeneous velocity field ${\bm U}$ is formally considered,
since it is convenient for later purposes. Its presence does not
affect the value of the integral on the right hand side of Eq.\
(\ref{4.3d}). Note that since the HCS is normal, i.e. all the time
dependence occurs through the temperature, the time derivative is
replaced by the scaling operator for cooling, $-\zeta _{0}\left(
n,T\right) T\partial/
\partial T$, of Eq. (\ref{2.22}). Equations (\ref{4.3d}) and
(\ref{4.4}) constitute a pair of time independent equations to
determine both $\rho _{h}(\Gamma ;n,T, \mathbf{U})$ and $\zeta
_{0}\left( n,T\right) $. Once the latter is known, $T_{h}(t)$ is
determined from the solution to Eq.\ (\ref{4.3a}) which becomes
\begin{equation}
\partial _{t}T_{h}(t)=-\zeta _{0}\left[ n_{h},T_{h}(t)\right] T_{h}(t).
\label{4.5}
\end{equation}
Finally, $\rho _{h}(\Gamma ;n,T,{\bm U})$ is also evaluated at
$n=n_{h}$, $ T=T_{h}(t) $, and ${\bm U}={\bm U}_{h}={\bm 0}$.

This definition of the HCS in terms of the solution to the Liouville
equation provides the first of the unknown hydrodynamic parameters,
the zeroth order cooling rate $\zeta _{0}\left[
n_{h},T_{h}(t)\right] $. It now has a precise and exact definition,
Eq. (\ref{4.3d}), from which to determine its density and
temperature dependence.

It is easily verified that the equilibrium probability densities for
normal fluids (e.g., Gibbs ensembles or maximum entropy ensembles
for conserved densities) are not solutions to Eq.\ (\ref{4.4}).
Thus, the effects of inelastic collisions are two-fold. First, they
introduce an inherent time dependence due to energy loss, through
$T_{h}(t)$, and second, they change the form of the probability
density in a way that cannot be simply represented by the global
invariants for a normal fluid. For the analysis here, it is assumed
that  Eq.\ (\ref{4.4}) can be solved to good approximation and its
solution is taken as known. The existence of this solution is
supported by molecular dynamics simulations that show a rapid
approach to a state whose granular temperature obeys Eq.\
(\ref{4.5}), with uniform density and flow velocity \cite{HCSMD}.
Furthermore, at sufficiently low density, the reduced distribution
function obtained by integrating out all degrees of freedom but one
in Eqs.\ (\ref{4.3d}) and (\ref{4.4}), should be approximated by the
corresponding Boltzmann limit. This limit also supports a HCS
solution, as verified by simulations of the Boltzmann equation using
the direct simulation Monte Carlo method \cite{HCSBoltz}.

\subsection{A Class of Homogeneous Solutions to the Liouville Equation}

Since cooling is inherent to the inelastic collisions for all
states, it is useful to consider the time dependence of a general
state  comprised of that through $T_{h}(t)$ plus any residual time
dependence, i.e., $\rho \left( \Gamma ;t\right) =\rho \left[ \Gamma
;n_{h},T_{h}(t),{\bm U};t\right] $. Then the Liouville equation
(\ref{3.6}) takes the form
\begin{equation}
\left( \partial _{t}+\overline{\mathcal{L}}_{T}\right) \rho \left(
\Gamma ;n,T,{\bm U};t\right) =0.  \label{4.5a}
\end{equation}
It is again understood that $t$ and $T$ are now independent
variables and the time derivative is taken at constant $T$. The new
generator for the dynamics $ \overline{\mathcal{L}}_{T}$, given in
Eq.\ (\ref{4.4a}), incorporates both that for the trajectories in
phase space and the effects of cooling. The final solution to this
equation is to be evaluated at $n=n_{h}$, $T= T_{h}(t)$. In this
representation, the distribution function of the HCS, $\rho
_{h}(\Gamma ;n,T,{\bm U})$, is seen to be a stationary solution to
the Liouville equation, as shown by Eq.\ (\ref{4.4}).

Once $\rho _{h}(\Gamma ;n,T,{\bm U})$ is known, a class of other
solutions can be constructed for homogeneous perturbations of the
HCS. These are the microscopic analogues of the hydrodynamic
solution defined by $\widetilde{C}^{hyd}(n,T;{\bm 0},t)$ given in
Eq.\ (\ref{2.23a}). To see this consider the initial condition
\begin{eqnarray}
\rho \left( \Gamma ;0\right) &=&\rho _{h}\left( \Gamma
;n+\delta n,T+\delta T,{\bm U}+ \delta {\bm U}\right)  \nonumber \\
& \simeq &\rho _{h}\left( \Gamma ;n,T,{\bm U}\right) +\sum_{\alpha}
\Psi _{\alpha }\left( \Gamma ;n,T,{\bm U}\right) \delta y_{\alpha
}\left( 0\right) , \label{4.6}
\end{eqnarray}
with the spatially uniform perturbations $\{ \delta y_{\alpha} (0)
\} \equiv \left\{ \delta n,\delta T,\delta {\bm U}\right\} $ and
\begin{equation}
\Psi _{\alpha }\left( \Gamma ;n,T,{\bm U}\right) \equiv \left(
\frac{\partial \rho _{h}\left( \Gamma ;n,T,{\bm U}\right)
}{\partial y_{\alpha }}\right)_{y_{\beta }\neq y_{\alpha }}.
\label{4.7}
\end{equation}
The corresponding solution to the Liouville equation (\ref{4.5a}) is
\begin{eqnarray}
\rho \left( \Gamma, t\right) &=&e^{-\overline{\mathcal{L}}
_{T}t}\left[ \rho _{h}\left( \Gamma ;n,T,{\bm U}\right)
+\sum_{\alpha} \Psi _{\alpha }\left( \Gamma ;n,T,{\bm U}\right)
\delta y_{\alpha }\left( 0\right)
\right]  \nonumber \\
&=&\rho _{h}\left( \Gamma ;n,T,{\bm U}\right)
+e^{-\overline{\mathcal{L}} _{T}t}\sum_{\alpha} \Psi _{\alpha
}\left( \Gamma ;n,T,{\bm U}\right) \delta y_{\alpha }\left(
0\right). \label{4.8}
\end{eqnarray}
The second equality follows from the stationarity of $\rho _{h}$ for
the generator $\overline{\mathcal{L}}_{T}$, Eq.\ (\ref{4.4}). The
second term in the last expression can be evaluated using the
identity
\begin{equation}
\overline{\mathcal{L}}_{T}\Psi _{\alpha }\left( \Gamma ;n,T,{\bm
U} \right) =\sum_{\beta} \Psi _{\beta }\left( \Gamma ;n,T,{\bm
U}\right) \mathcal{K} _{\beta \alpha }^{hyd}(n,T;{\bm 0}),
\label{4.10}
\end{equation}
which follows by direct calculation using the definition of $\overline{%
\mathcal{L}}_{T}$ and $\Psi _{\alpha }$. Here
$\mathcal{K}^{hyd}(n,T;{\bm 0})$ is the same matrix as obtained by
evaluating Eqs.\ (\ref{2.14})-(\ref{2.15}) at ${\bm k}={\bm 0}$. The
proof is given in Appendix \ref{ap3}, where it is also shown that
Eq. (\ref{4.10}) in turn gives
\begin{equation}
e^{-\overline{\mathcal{L}}_{T}t}\Psi _{\alpha }\left( \Gamma
;n,T,{\bm U} \right) =\sum_{\beta} \Psi _{\beta }\left( \Gamma
;n,T,{\bm U}\right) \widetilde{C} _{\beta \alpha }^{hyd}\left(
n,T;{\bm 0},t\right).  \label{4.11}
\end{equation}
Here $\widetilde{C}_{\beta \alpha }^{hyd}\left( n,T;{\bm 0},t\right)
$ is the same hydrodynamic response matrix of Eq.\ (\ref{2.23a}).
The exact solution to the Liouville equation is therefore
\begin{equation}
\rho \left( \Gamma ;t\right) =\rho _{h}\left( \Gamma ;n,T,%
{\bm U}\right) + \sum_{\alpha}\Psi _{\alpha }\left( \Gamma
;n,T,{\bm U}\right) \delta y_{\alpha }\left( t\right),
\label{4.12}
\end{equation}
where
\begin{equation}
\delta y_{\alpha }\left(t \right) = \sum_{\beta}
\widetilde{C}_{\alpha \beta }^{hyd}\left( n,T;{\bm 0},t\right)
\delta y_{\beta }\left(0 \right) .  \label{4.12a}
\end{equation}
The special choice of the $\Psi _{\alpha }$ for initial
perturbations is seen to excite only hydrodynamic modes, and no
other microscopic homogeneous excitations. In this sense, the $\Psi
_{\alpha }\left( \Gamma ;n,T,{\bm U} \right) $ can be considered the
microscopic hydrodynamic modes in the long wavelength limit. For a
normal fluid, they become functions of the global invariants and
therefore time independent, which are indeed the hydrodynamic
excitation at $k=0$ in that case. For a granular fluid, as discussed
in the previous section, there is a nontrivial dynamics even at
$k=0$. The analogy can be made more direct by rewriting Eq.\
(\ref{4.11}) in the form
\begin{equation}
\sum_{\beta} \mathcal{U}_{\alpha \beta }\left( t,T\right) \Psi
_{\beta }\left( \Gamma ;n,T,{\bm U}\right) =\Psi _{\alpha }\left(
\Gamma ;n,T,{\bm U}\right) , \label{4.13}
\end{equation}
where the new matrix evolution operator $\mathcal{U}\left(
t,T\right) $ is defined as
\begin{equation}
\mathcal{U}_{\alpha \beta }\left( t,T\right) \equiv  \widetilde{C}
^{hyd -1}_{\beta \alpha} \left( n,T;{\bm 0},t\right)  e^{-
\overline{\mathcal{L}}_{T}t}.  \label{4.14}
\end{equation}
The dynamics described by $\mathcal{U}\left( t,T\right) $ is the
Liouville dynamics, compensated for both effects of cooling,
through the scaling generator in $\overline{\mathcal{L}}_{T}$, and
the dynamics of homogeneous perturbations, through the response
function $\widetilde{C} ^{hyd}\left( n,T;{\bm 0},t\right) $.
Consequently, $\mathcal{U}\left( t,T\right) $ provides the
dynamics associated with spatial perturbations. It will be seen
below that this operator defines the time dependence of the
correlation functions representing all transport coefficients.
Equation (\ref{4.13}) shows that the functions $\{ \Psi _{\alpha }
\} $ are the global invariants for this dynamics. Note that
although it is convenient to keep ${\bm U} \neq {\bm 0}$ at a
formal level for the discussion here, there is no problem in
taking the limit ${\bm U} \rightarrow {\bm 0}$ in all the obtained
results.

\section{Initial preparation and linear response}
\label{sec4}

In this section, the response of the system to an initial spatial
perturbation in the hydrodynamic fields relative to the HCS is
studied, in order to extract the exact analogue of the hydrodynamic
transport matrix described on phenomenological grounds in Eqs.\
(\ref{2.14})-(\ref{2.15}) above. Consider an initial perturbation
generalizing (\ref{4.6}) to inhomogeneous states,
\begin{equation}
\rho \left( \Gamma ;0\right) =\rho _{h}\left( \Gamma ;n,T \right) +
\sum_{\alpha} \int d{\bm r}\, \psi _{\alpha }\left( \Gamma ;n,T;
{\bm r}\right) \delta y_{\alpha }\left( {\bm r},0\right) ,
\label{6.0}
\end{equation}
where now $\left\{ \delta y_{\alpha }({\bm r},0)\right\} = \left\{
\delta n({\bm r},0),\delta T({\bm r},0),\delta {\bm U}( {\bm
r},0)\right\} $ are the initial space dependent deviations of the
hydrodynamic fields from their values in the reference HCS  and we
have taken for simplicity ${\bm U}=0$. Thus the argument ${\bm U}$
will be omitted in the following when it vanishes. The $\left\{ \psi
_{\alpha }\right\} $ are spatial densities corresponding to the
invariants $\left\{ \Psi _{\alpha }\right\} $,
\begin{equation}
\Psi _{\alpha }\left( \Gamma ;n,T\right) =\int d{\bm r}\, \psi
_{\alpha }\left( \Gamma ;n,T;{\bm r}\right) . \label{6.01}
\end{equation}
This choice of this initial condition assures that the long
wavelength limit of the perturbation gives the hydrodynamic solution
(\ref{4.12}). The hydrodynamic fields $\left\{ \delta y_{\alpha
}({\bm r},t)\right\} $ as identified below Eq.\ (\ref{6.0}),  are
averages of corresponding phase functions $\left\{ a_{\alpha }\left(
\Gamma ;n,T;{\bm r}\right) \right\} $, or, more precisely, it is
\begin{equation}
\delta y_{\alpha }\left({\bm r},t\right) =\int d\Gamma\,  \left[\rho
\left( \Gamma; t\right) - \rho_{h}(\Gamma;n,T) \right]\ a_{\alpha
}\left( \Gamma ;n,T;{\bm r}\right) ,  \label{6.02}
\end{equation}
\begin{equation}
\left\{ a_{\alpha }\left( \Gamma ;n,T;{\bm r} \right) \right\}
\equiv \left\{ \mathcal{N} \left( \Gamma ;{\bm r} \right) ,
\frac{1}{e_{0,T}}\left[ \mathcal{E}\left( \Gamma ; {\bm r} \right)
-e_{0,n} \mathcal{N} \left( \Gamma ;{\bm r}\right) \right]
,\frac{\bm{\mathcal{G}}(\Gamma;{\bm r})}{nm} \right\} . \label{6.03}
\end{equation}
The local microscopic phase functions for the number density,
$\mathcal{N} \left( {\bm r} \right) $, energy density, $\mathcal{E}
\left( {\bm r} \right) $, and momentum density, $\bm{\mathcal{G}}
\left( {\bm r} \right)$, are
\begin{equation}
\left(
\begin{array}{c}
\mathcal{N}\left( \Gamma ;{\bm r}\right) \\
\mathcal{E}\left( \Gamma ; {\bm r} \right) \\
\bm{\mathcal{G}}\left( \Gamma ; {\bm r} \right)
\end{array}
\right)  \equiv \sum_{r=1}^{N} \delta \left( {\bm r}-{\bm q}_{r}
\right) \left(
\begin{array}{c}
1 \\
\frac{m v_{r}^{2}}{2}+\frac{1}{2}\sum_{r\neq s}V\left( {\bm
q}_{rs}\right)
\\
m {\bm v}_{r}
\end{array}
 \right) .  \label{6.1}
\end{equation}
In this expression, ${\bm q}_{rs}={\bm q}_{r}-{\bm q}_{s}$ and
$V\left( {\bm q}_{rs}\right) $ is the pair potential for the
conservative part of the interaction among particles, as discussed
in Appendix \ref{ap2}. Normalization of both $\rho \left( \Gamma
;0\right) $ and $\rho _{h}\left( \Gamma ;n,T\right) $, and the
representation (\ref{6.02}) for $\delta y_{\alpha }\left(
\mathbf{r},0\right) $ require
\begin{equation}
\int d\Gamma\,  \psi _{\alpha }\left( \Gamma ;n, T; {\bm r}
\right) =0,\quad  \int d\Gamma\,  a_{\alpha }\left( \Gamma
;n,T;{\bm r}\right) \psi _{\beta }\left( \Gamma ;n,T;{\bm
r}^{\prime }\right) =\delta _{\alpha \beta }\delta \left( {\bm
r}-{\bm r}^{\prime }\right) .  \label{6.2}
\end{equation}
The second relation above shows that $\left\{ a_{\beta }\right\} $
and $\left\{ \psi _{\beta }\right\} $ comprise two biorthogonal
sets.

The conditions (\ref{6.01}) and (\ref{6.2}) are satisfied if the
densities $ \psi _{\alpha }\left( \Gamma ;{\bm r}\right) $ are
generated from a \emph{ local} HCS distribution, $\rho _{lh}\left(
\Gamma \mid \left\{ y_{\beta }\right\} \right) $, through
\begin{equation}
\psi _{\alpha }\left( \Gamma ;n,T ;{\bm r} \right) = \left[
\frac{\delta \rho _{lh}\left( \Gamma | \left\{ y_{\beta }\right\}
\right) }{\delta y_{\alpha }\left( {\bm r} \right) }\right]_{
\{y_{\beta} \}=\{ n,T,{\bm 0}\}}. \label{6.4c}
\end{equation}
This local HCS distribution is the analogue of the local
equilibrium \ distribution for a normal fluid. Qualitatively, it
corresponds to the condition that each local cell has an HCS
distribution characterized by the parameters $\left\{ y_{\alpha
}\left( {\bm r},0\right) =y_{\alpha, h }+\delta y_{\alpha }\left(
{\bm r},0\right) \right\} $. This characterization as a local form
for the HCS requires
\begin{equation}
\delta y_{\alpha }\left( {\bm r},0\right) =\int d\Gamma\,  \left[
\rho _{lh}\left( \Gamma \mid \left\{ y_{\beta }(0)\right\} \right) -
\rho_{h} ( \Gamma,n,T) \right] a_{\alpha }\left( \Gamma ;\left\{
y_{\beta }\right\} ;{\bm r}\right), \label{6.4d}
\end{equation}
\begin{equation}
\rho _{l h}\left( \Gamma | \{y_{\alpha,h} \} \right) =\rho
_{h}\left( \Gamma ; \{ y_{\alpha,h} \}\right), \label{6.3}
\end{equation}
\begin{equation}
\int d{\bm r}_{1} \ldots \int d{\bm r}_{n}\, \left[ \frac{\delta
^{n}\rho _{lh}}{ \delta y_{\alpha }\left( {\bm r}_{1}\right)
 \ldots \delta y_{\beta }\left(
{\bm r}_{n}\right) }\right]_{\{y_{\beta}\}=\{n,T,{\bm
U}\}}=\frac{\partial ^{n}\rho _{h}(\Gamma; n,T,{\bm U})}{\partial
y_{\alpha, h}\ldots
\partial y_{\beta, h }}. \label{6.4}
\end{equation}
The first equality states that the local state has the exact
average values for the $\left\{ a_{\alpha }\right\} $; the other
two refer to the uniform limit. This is sufficient for the
conditions (\ref{6.01}) and (\ref{6.2}) to be verified.

The formal solution to the Liouville equation for this initial
condition, generalizing Eq.\ (\ref{4.8}) to spatially varying
perturbations, is
\begin{equation}
\rho \left( \Gamma ;t\right) =\rho _{h}\left( \Gamma ;n,T\right) +
\sum_{\alpha} \int d{\bm r}\, e^{-\overline{\mathcal{L}}_{T}t}\psi
_{\alpha }\left( \Gamma ;n,T;{\bm r}\right) \delta y_{\alpha }\left(
{\bm r},0\right) .  \label{6.5d}
\end{equation}
The response in the hydrodynamic fields is therefore, in matrix
notation,
\begin{equation}
\delta y\left( {\bm r},t\right) =\int d{\bm r}^{\prime }\,
C\left(n,T; {\bm r}-{\bm r}^{\prime },t\right) \delta y\left( {\bm
r}^{\prime},0\right) , \label{6.7}
\end{equation}
with the matrix of response functions defined by
\begin{equation}
C_{\alpha \beta }\left( n,T;{\bm r}-{\bm r}^{\prime },t\right) =
\int d\Gamma\,  a_{\alpha }\left( \Gamma ;n,T;{\bm r}\right) e^{-\overline{%
\mathcal{L}}_{T}t}\psi _{\beta }\left( \Gamma ;n,T,{\bm r}^{\prime
}\right).  \label{6.7b}
\end{equation}
This notation expresses the translational invariance of the
reference HCS and the generator for the dynamics. This is the
exact response function for the chosen initial perturbation,
representing all possible excitations of the many-body dynamics.
The corresponding Fourier representation is
\begin{equation}
\widetilde{C}_{\alpha \beta }\left( n,T;{\bm k},t\right) = V^{-1}
\int d\Gamma\,  \widetilde{a}_{\alpha }\left( \Gamma ;n,T;{\bm
k}\right) e^{- \overline{\mathcal{L}}_{T}t}\widetilde{\psi }_{\beta
}(\Gamma ;n,T,-{\bm k }).  \label{6.8a}
\end{equation}
These response functions will be the primary objects of study in all of the
following.

As a consequence of Eq.\ (\ref{6.01}), at ${\bm k=0}$ the
$\widetilde{\psi }_{\alpha }(\Gamma ;n,T,{\bm k})$'s become the
invariants
\begin{equation}
\widetilde{\psi }_{\alpha }(\Gamma ;n,T;{\bm 0})=\int d{\bm r}\,
\psi _{\alpha }\left( \Gamma ;n,T;{\bm r}\right) =\Psi _{\alpha
}\left( \Gamma ;n,T\right) .  \label{6.8b}
\end{equation}
Therefore, it follows from Eqs.\ (\ref{4.11}) and (\ref{6.2}) that
the exact response function in the long wavelength limit is the
same as that from hydrodynamics,
\begin{equation}
\widetilde{C}_{\alpha \beta }\left( n,T;{\bm 0},t\right)
=\widetilde{C} _{\alpha \beta }^{hyd}\left( n,T;{\bm 0},t\right) ,
\label{6.8c}
\end{equation}
with $\widetilde{C}_{\alpha \beta }^{hyd}(n,T;{\bm 0},t)$ given by
Eq.\ (\ref{2.23a}). By construction, the choice of initial
preparation made has eliminated all microscopic homogeneous
transients and only the hydrodynamic mode is excited. This result is
exact, and shows that the microscopic dynamics supports a
hydrodynamic response at long wavelengths. At this formal level,
analyticity in $k$ is sufficient to admit the possibility of
Navier-Stokes modes.

To identify the macroscopic hydrodynamic equations, it is useful
first to write an exact equation for the response function in a form
similar to (\ref {2.21}),
\begin{equation}
\left[ \partial _{t}-\zeta _{0}\left( n,T\right) T
\frac{\partial}{\partial T}+\mathcal{K} \left( n,T;{\bm k},t\right)
\right] \widetilde{C}\left( n,T;{\bm k} ,t\right)
=0,\hspace{0.3in}\widetilde{C}_{\alpha \beta }\left( n,T;{\bm k}
,0\right) =\delta _{\alpha \beta },  \label{6.8}
\end{equation}
which provides a formal expression for the generalized transport
matrix,
\begin{eqnarray}
\mathcal{K}\left( n,T;{\bm k},t\right) &=&-\left\{ \left[ \partial
_{t}-\zeta _{0}\left( n,T\right) T \frac{\partial}{\partial T}
\right] \widetilde{C}\left( n,T;{\bm k},t\right) \right\}
\widetilde{C}^{-1}\left( n,T;{\bm k}
,t\right)  \nonumber \\
&=&\mathcal{K}^{hyd}\left( n,T;{\bm 0}\right)  \nonumber \\
&&-\left\{ \left[ \partial _{t}-\zeta _{0}\left( n,T\right) T
\frac{\partial}{\partial T}+ \mathcal{K}^{hyd}\left( n,T; {\bm
0}\right) \right] \widetilde{C}\left( n,T;{\bm k},t\right) \right\}
\widetilde{C}^{-1}\left( n,T;{\bm k} ,t\right) . \nonumber \\
\label{6.10a}
\end{eqnarray}
The contribution from ${\bm k}={\bm 0}$  has been extracted
explicitly, since Eq.\ (\ref{6.8c}) implies it is exact at all
times. The other term is proportional to the time derivative
relevant for the homogeneous dynamics and therefore is of order $
k $. The full hydrodynamic matrix, as given by Eqs.\
(\ref{2.14})-(\ref{2.15}), when it exists, follows from this
formal result for small $k$ (long wavelengths) and long times,
\begin{equation}
\mathcal{K}^{hyd}\left( n,T;{\bm k}\right) \equiv \lim_{t>>t_{0},k<<k_{0}}%
\mathcal{K}\left( n,T;{\bm k},t\right).  \label{6.10d}
\end{equation}
The characteristic time $t_{0}$ and wavelength $k_{0}^{-1}$ are
expected to be the mean free time and mean free path, respectively.
Comparison of this expression with the form (\ref{2.14}) provides a
``derivation'' of the linear hydrodynamic equations, and also gives
the coefficients of those equations in terms of the response
functions. A detailed comparison up through order $k^{2}$ is the
objective of the next few sections.

The result (\ref{6.10a}) is the first of three exact representations
of the transport matrix to be obtained here. Its expansion to order
$k^{2}$ leads directly to the Einstein-Helfand representation of the
transport coefficients as the long time limit of time derivatives of
correlation functions. This is analogous to the diffusion
coefficient $D$ represented in terms of the time derivative of the
mean square displacement, i.e. the first representation in Eq.\
(\ref{1.5}) \cite{DBB05}. However, due to the homogeneous state
dynamics, the relevant time derivative is $ \left[
\partial _{t}-\zeta _{0}\left( n,T\right) T \partial / \partial T
+\mathcal{K}^{hyd}\left( n,T;{\bm 0}\right) \right], $ so this form
may not be optimal in practice. There is an intermediate Helfand
form which entails correlation functions with non-zero long time
limits determining the transport coefficients. Finally, the third
equivalent representation is the Green-Kubo form in terms of time
integrals of correlation functions. These second and third forms are
given in the next section and utilized to implement the $k$
expansion.

\section{Navier-Stokes Hydrodynamics}
\label{sec5}

The linearized Navier-Stokes equations follow from an evaluation of
the transport matrix of Eq. (\ref{6.10a}) to order $k^{2}$. This can
be accomplished by a direct expansion of $\widetilde{C}_{\alpha
\beta }\left( n,T;{\bm k},t\right) $ in powers of $k$ \cite{DBB05}.
It is somewhat more instructive to proceed in a different manner,
using the microscopic conservation laws to expose the dominant $k$
dependence. This allows interpretation of the phase functions
occurring in the correlation functions of the final expressions.

\subsection{Consequences of Conservation Laws}

For normal fluids, the variables $\widetilde{a}_{\alpha }\left(
\Gamma ;n,T; {\bm k}\right) $ are the Fourier transforms of linear
combinations of the local conserved densities, so their time
derivatives are equal to $i{\bm k} \cdot {\bm f}_{\alpha }\left(
\Gamma ;n,T;{\bm k}\right) $, where the ${\bm f}_{\alpha }$ are the
associated microscopic fluxes. The proportionality to $k$ of the
time derivatives means that they vanish in the long wavelength
limit, as appropriate for a conserved density. This allows
evaluation of the time derivative in Eq.\ (\ref{6.10a}) and shows
the transport matrix is of order $k$. Then shifting the time
dependence to the other density in the response function and using
again the conservation law, the dependence through order $k^{2}$ is
exposed in terms of correlation functions involving the fluxes
\cite{McL}. It is somewhat more complicated for granular fluids,
although the general idea is the same.

Consider first the time derivative occurring in Eq.\ (\ref{6.10a}).
Using the definition of the matrix of response functions, Eq.\
(\ref{6.8a}), it can be transformed into
\begin{equation}
\left[ \partial _{t}- \zeta _{0}\left( n,T\right) T
\frac{\partial}{\partial T}+ \mathcal{K}^{hyd}\left( n,T;{\bm
0}\right) \right] \widetilde{C}\left( n,T;{\bm k},t\right) \nonumber
\end{equation}
\begin{equation}  = V^{-1} \int
d\Gamma\, \left\{ \left[ L-\zeta_{0}\left( n,T\right) T
\frac{\partial}{\partial T}+\mathcal{K}^{hyd}\left( n,T; {\bm
0}\right) \right] \widetilde{a}\left( \Gamma ;n,T;{\bm k}\right)
\right\} e^{-\overline{ \mathcal{L}}_{T}t}\widetilde{\psi }(\Gamma
;n,T;- {\bm k}), \label{5.1}
\end{equation}
where the adjoint generator $L$ of $\overline{L}$  has been
introduced. As mentioned above, for normal fluids it is
$L\widetilde{a}_{\alpha}= i{\bm k} \cdot {\bm f}_{\alpha }\left(
\Gamma ;n,T;{\bm k}\right)$. Here, the inelastic collisions give
rise to an additional energy loss $\widetilde{w}(\Gamma ;{\bm
k})$, which is not proportional to $k$ and therefore cannot be
absorbed in the flux. The new relationships are (see Appendix
\ref{ap4}):
\begin{equation}
L\widetilde{a}_{\alpha }(\Gamma;n,T;{\bm k})=i {\bm k} \cdot {\bm
f}_{\alpha } \left( \Gamma ;n,T;{\bm k} \right) -\delta _{\alpha
2}\, \frac{\widetilde{w}(\Gamma ;{\bm k} )}{e_{0,T}}. \label{5.2}
\end{equation}
The detailed expressions of ${\bm f}_{\alpha } \left( \Gamma
;n,T;{\bm k} \right)$ and  $\widetilde{w}(\Gamma ;{\bm k} )$ are
given in Appendix \ref{ap4}. The pre-factor $1/e_{0,T}$ in the
energy loss term, appears because of the definition of
$\widetilde{a}_{2}$. For $\alpha \neq 2$, the right side gives the
usual fluxes for number and momentum density. Inclusion of the
additional terms $-\zeta _{0}\left( n,T\right) T\partial/ \partial
T+\mathcal{K} ^{hyd}\left( n,T;{\bm 0}\right) $ in Eq.\
(\ref{5.1}) modifies this result to
\begin{equation}
\left[ L-\zeta _{0}\left( n,T\right) T \frac{\partial}{\partial T}
\right] \widetilde{a}_{\alpha }(\Gamma; n,T;{\bm k}) +
\sum_{\beta} \mathcal{K}^{hyd}_{\alpha \beta} \left( n,T;{\bm
0}\right) \widetilde{a}_{\beta }(\Gamma; n,T;{\bm k}) \nonumber
\end{equation}
\begin{equation}
=i {\bm k} \cdot {\bm f}_{\alpha } \left( \Gamma ;n,T;{\bm
k}\right) -\delta _{\alpha 2}\widetilde{\ell }(\Gamma ;n,T;{\bm
k}), \label{5.3}
\end{equation}
where $\widetilde{\ell }(\Gamma ;n,T; {\bm k})$ is defined by
\begin{equation}
\widetilde{\ell }(\Gamma ;n,T;{\bm k}) \equiv
\frac{1}{e_{0,T}}\left[ \widetilde{w}(\Gamma ;{\bm k})- V^{-1}
\sum_{\alpha} \widetilde{a}_{\alpha } \left( \Gamma ;n,T; {\bm
k}\right) \int d\Gamma\,  \Psi _{\alpha }\left( \Gamma ;n,T\right)
\widetilde{w}(\Gamma ;{\bm 0})\right] . \label{5.4}
\end{equation}
The first contribution in this expression is the phase function
whose average in the HCS gives the cooling rate,
\begin{equation}
\frac{1}{e_{0,T}} \int d\Gamma\,  \rho _{h}\left( \Gamma ;n,T\right)
\widetilde{w }(\Gamma ;{\bm 0})=\zeta _{0}(n,T)T. \label{5.4c}
\end{equation}
The remaining terms assure that $\widetilde{\ell }(\Gamma
;n,T;{\bm 0})$ is orthogonal to the invariants, namely that
\begin{equation}
\widetilde{\ell }(\Gamma ;n,T;{\bm 0})=\left( 1-P^{\dagger
}\right) \frac{\widetilde{w}(\Gamma ;{\bm 0}) }{e_{0,T}}= - \left(
1-P^{\dagger }\right) \frac{LE(\Gamma)}{e_{0,T}}. \label{5.4b}
\end{equation}
Here, $P^{\dagger }$ is the projection operator onto the set
$\left\{ \widetilde{a} _{\alpha }\left( \Gamma ;n,T,{\bm 0}\right)
\right\} $,
\begin{equation}
P^{\dagger }X (\Gamma) = V^{-1} \sum_{\alpha}
\widetilde{a}_{\alpha }\left( \Gamma ;n,T;{\bm 0}\right) \int
d\Gamma \Psi _{\alpha }\left( \Gamma ;n,T\right) X(\Gamma).
\label{5.4a}
\end{equation}
To verify that $P^{\dagger}$ is really a projection operator,
recall that as a consequence of Eq.\ (\ref{6.2}), $\left\{
\widetilde{a}_{\alpha }\left( \Gamma ;n,T,;{\bm 0}\right) \right\}
$ and $\left\{ \Psi _{\alpha }\left( \Gamma ;n,T \right) \right\}
$ form a biorthogonal set,
\begin{equation}
V^{-1} \int d\Gamma\, \widetilde{a}_{\alpha} (\Gamma;n,T;{\bm 0})
\Psi_{\beta}(\Gamma;n,T) =\delta_{\alpha \beta}.
\end{equation}
Use of Eq.\ (\ref{5.3}) in Eq.\  (\ref{5.1}) shows that the
correlation functions $\widetilde{C} _{\alpha \beta }\left( n,T;{\bm
k},t\right) $ obey the equations
\begin{equation}
\left[ \partial _{t}-\zeta _{0}T \frac{\partial}{\partial
T}+\mathcal{K}^{hyd}\left( n,T; {\bm 0}\right) \right]
\widetilde{C}\left( n,T;{\bm k},t\right) =i{\bm k} \cdot
\widetilde{\bm D}\left( n,T;{\bm k},t\right) - \widetilde{S} \left(
n,T;{\bm k},t\right) .  \label{5.4d}
\end{equation}
The new correlation functions on the right hand side,
$\widetilde{\bm D}_{\alpha \beta}\left( n,T;{\bm k},t\right)$ and
$\widetilde{S}_{\alpha  \beta} \left( n,T;{\bm k},t\right)$, are
similar to $\widetilde{C}_{\alpha \beta} \left( n,T;{\bm k},t\right)
$ but with $\widetilde{a}_{\alpha }$ replaced by $\widetilde{\bm
f}_{\alpha }$ and $ \delta_{\alpha 2} \widetilde{\ell }$,
respectively,
\begin{equation}
\widetilde{\bm D}_{\alpha \beta }\left( n,T;{\bm k},t\right)
=V^{-1}\int d\Gamma\,  \widetilde{\bm f}_{\alpha }(\Gamma
;n,T;{\bm k})e^{-\overline{ \mathcal{L}}_{T}t}\widetilde{\psi
}_{\beta }(\Gamma ;n,T;-{\bm k}), \label{5.4e}
\end{equation}
\begin{equation}
\widetilde{S}_{\alpha \beta }\left( n,T;{\bm k},t\right) =
\delta_{\alpha 2} V^{-1}\int d\Gamma\,  \widetilde{\ell }(\Gamma
;n,T;{\bm k})e^{-\overline{\mathcal{L}} _{T}t}\widetilde{\psi
}_{\beta }(\Gamma ;n,T;-{\bm k}).  \label{5.4f}
\end{equation}
The utility of Eq.\ (\ref{5.4d}) is that its use in Eq.\
(\ref{6.10a}) leads to an expression in which the  transport matrix
is exposed exactly through first order in $k$,
\begin{equation}
\mathcal{K}\left( n,T;{\bm k},t\right) =\mathcal{K}^{hyd}\left(
n,T; {\bm 0}\right) -\left[ i {\bm k} \cdot \widetilde{\bm
D}\left( n,T;{\bm k},t\right) - \widetilde{S}\left( n,T;{\bm
k},t\right) \right] \widetilde{C}^{-1}\left( n,T;{\bm k},t\right)
. \label{5.5}
\end{equation}
It follows from Eqs.\, ({\ref{4.11}) and (\ref{5.4b}) that
$\widetilde{S}\left( n,T;{\bm 0} ,t\right) =0$, so the term between
square brackets on the right hand side of Eq.\ (\ref{5.5}) is at
least of order $k$, as said above. However, this representation is
still not optimal since the right hand side has the homogeneous
dynamics of $\widetilde{C}\left( n,T; {\bm 0},t\right) $ that should
be cancelled. This technical point is addressed by incorporating
$\widetilde{C}^{-1}\left( n,T;{\bm 0},t\right) $ in the evolution
operator for the correlation functions by using $\mathcal{U}
_{\alpha \beta }\left( t,T\right) $ introduced in Eq.\ (\ref{4.14}),
i.e. defining
\begin{equation}
\widetilde{\psi }_{\alpha }(\Gamma ;n,T;{\bm k},t)\equiv
\sum_{\beta} \mathcal{U} _{\alpha \beta }\left( t,T\right)
\widetilde{\psi }_{\beta }(\Gamma ;n,T; {\bm k}). \label{5.5aaa}
\end{equation}
There is no ${\bm k=0}$ dynamics for $\widetilde{\psi }_{\alpha
}(\Gamma ;n,T;{\bm k},t)$ since $\widetilde{\psi }_{\alpha
}(\Gamma ;n,T;{\bm  0})$ is an invariant; see Eq.\ (\ref{6.8b}).
The transport matrix (\ref{5.5}) then becomes
\begin{equation}
\mathcal{K}(n,T;{\bm k,}t)=\mathcal{K}^{hyd}\left( n,T;{\bm
0}\right)-\left[ i {\bm k} \cdot \overline{\bm D}(n,T;{\bm
k},t)-\overline{S}(n,T;{\bm k},t)\right] \overline{C}^{-1}(n,T;{\bm
k},t). \label{5.5c}
\end{equation}
The correlation functions with the over-bar are the same as those
with the tilde, except that now are defined with the dynamics of
(\ref{5.5aaa}),
\begin{equation}
\overline{C}_{\alpha \beta }(n,T;{\bm k},t)=V^{-1}\int d\Gamma\,
\widetilde{a}_{\alpha }(\Gamma ;n,T;{\bm k})\widetilde{\psi
}_{\beta }(\Gamma ;n,T;- {\bm k},t),  \label{5.5d}
\end{equation}
\begin{equation}
\overline{\bm D}_{\alpha \beta }(n,T;{\bm k},t)=V^{-1}\int
d\Gamma\, \widetilde{\bm f}_{\alpha }(\Gamma ;n,T;{\bm
k})\widetilde{\psi }_{\beta }(\Gamma ;n,T;-{\bm k},t),
\label{5.5e}
\end{equation}
\begin{equation}
\overline{S}_{\alpha \beta }(n,T;{\bm k},t)=\delta _{\alpha
2}V^{-1}\int d\Gamma\,  \widetilde{\ell }(\Gamma ;n,T;{\bm
k})\widetilde{\psi }_{\beta }(\Gamma ;n,T;-{\bm k},t).
\label{5.5f}
\end{equation}
Equation (\ref{5.5c}) gives the \emph{intermediate Helfand}
representation referred to at the end of the last section. It has
the advantage of being expressed in terms of the appropriate
dynamics of $\mathcal{U}_{\alpha \beta }\left( t,T\right) $, as
well as avoiding the complex time derivative of Eq.\
(\ref{6.10a}).

The equivalent Green-Kubo form is obtained by representing the
correlation functions in Eq.\ (\ref{5.5c}) as time integrals. This
is accomplished by observing that there are ``conjugate''
conservation laws associated with $ \widetilde{\psi }_{\alpha
}(\Gamma ;n,T;{\bm k},t)$. Their existence follows from the fact
that the $\psi_{\alpha }$'s are the densities associated with the
invariants $\Psi _{\alpha }$'s. The conjugate conservation laws are
\begin{equation}
\partial _{t}\widetilde{\psi }_{\alpha }(\Gamma ;n,T;{\bm
k},t)-i{\bm k} \cdot \widetilde{\bm \gamma}_{\alpha }\left( \Gamma
;n,T;{\bm k} ,t\right) =0.  \label{5.5aa}
\end{equation}
The new fluxes $\widetilde{\bm \gamma }_{\alpha }$ are identified in
Appendix \ref{ap4}  as
\begin{equation}
\widetilde{\bm \gamma }_{\alpha }\left( \Gamma ;n,T;{\bm k}
,t\right) =\sum_{\beta} \mathcal{U}_{\alpha \beta }\left(
t,T\right) \widetilde{\bm \gamma }_{\beta }\left( \Gamma ;n,T;{\bm
k}\right)  \label{5.5ab}
\end{equation}
with
\begin{equation}
i{\bm k} \cdot \widetilde{\bm \gamma }_{\alpha }\left( \Gamma ;n,T;
{\bm k}\right)  \equiv - \overline{\mathcal{L}}_{T}\widetilde{\psi
}_{\alpha }(\Gamma ;n,T;{\bm k})+\sum_{\beta} \mathcal{K}_{\beta
\alpha }^{hyd}(n,T;{\bm 0}) \widetilde{\psi }_{\beta }(\Gamma
;n,T;{\bm k}) .  \label{5.5ac}
\end{equation}
These new conservation laws give directly
\begin{equation}
\partial _{t}\overline{C}_{\alpha \beta }(n,T;{\bm k},t)+i{\bm k} \cdot
\overline{\bm E}_{\alpha \beta }(n,T;{\bm k},t)=0,  \label{5.6}
\end{equation}
\begin{equation}
\partial _{t}\overline{\bm D}_{\alpha \beta }(n,T;{\bm k},t)+i {\bm
k} \cdot  \overline{\sf F}_{\alpha \beta }(n,T;{\bm k},t)=0,
\label{5.6b}
\end{equation}
\begin{equation}
\partial _{t}\overline{S}_{\alpha \beta }(n,T;{\bm k,}t)+i {\bm k}
\cdot \overline{\bm N}_{\alpha \beta }(n,T;{\bm k},t)=0,
\label{5.6a}
\end{equation}
where
\begin{equation}
\overline{\bm E}_{\alpha \beta }(n,T;{\bm k},t)=V^{-1}\int d\Gamma\,
\widetilde{a}_{\alpha }(\Gamma ;n,T;{\bm k}) \widetilde{\bm
\gamma}_{\beta }(\Gamma ;n,T;-{\bm k},t), \label{5.6c}
\end{equation}
\begin{equation}
\overline{\sf F}_{\alpha \beta }(n,T;{\bm k},t)=V^{-1}\int d\Gamma\,
\widetilde{\bm f}_{\alpha }(\Gamma ;n,T;{\bm k}) \widetilde{\bm
\gamma } _{\beta }(\Gamma ;n,T;-{\bm k},t),  \label{5.6e}
\end{equation}
\begin{equation}
\overline{\bm N}_{\alpha \beta }(n,T;{\bm k},t)=\delta _{\alpha
2}V^{-1}\int d\Gamma\,  \widetilde{\ell }(\Gamma ;n,T;{\bm
k})\widetilde{\bm  \gamma }_{\beta }(\Gamma ;n,T;-{\bm k},t).
\label{5.6d}
\end{equation}
Note that ${\sf F}_{\alpha \beta}$ is a second-rank tensor.
Integrating Eqs.\ (\ref{5.6})-(\ref{5.6a}) allows
$\overline{C}(n,T;{\bm k},t)$, $\overline{\bm D}(n,T;{\bm k},t)$,
and $\overline{S}(n,T;{\bm k},t)$ to be eliminated from Eq.\
(\ref{5.5c}) in favor of $\overline{\bm E}$, $\overline{\sf F}$, and
$\overline{\bm N}$, exposing a higher order dependence on $k$,
\begin{eqnarray}
\mathcal{K}(n,T;{\bm k},t) & = & \mathcal{K}^{hyd}(n,T;{\bm 0})
-\left[ i {\bm k} \cdot \overline{\bm D}(n,T;{\bm k},0) \right.
\nonumber \\
& & \left. -\overline{S}(n,T;{\bm k},0)+ i {\bm k} \cdot
\int_{0}^{t} dt^{\prime}\, \overline{\bm N}(n,T;{\bm k},t^{\prime})+
{\bm k} {\bm k} : \int_{0}^{t} dt^{\prime}\, \overline{\sf
F}(n,T;{\bm
k},t^{\prime}) \right] \nonumber \\
& & \times \left[ I+ i {\bm k}  \cdot \int_{0}^{t} dt^{\prime}\,
\overline{C}^{-1}(n,T;{\bm k},t^{\prime}) \overline{\bm E}(n,T;{\bm
k},t^{\prime}) \overline{C}^{-1}(n,T;{\bm k},t^{\prime}) \right].
\label{5.6f}
\end{eqnarray}
This is the \emph{Green-Kubo} form for the transport matrix. An
advantage of this form is a further exposure of the explicit
dependence on $k$. In both  Eqs.\ (\ref{5.5c}) and (\ref{5.6f}),
relevant correlation functions are seen to be those composed from
the conserved densities $\left\{ \widetilde{a}_{\alpha },
\widetilde{\psi }_{\alpha }\right\} $, the fluxes of the two kinds
of conservation laws $\left\{ \widetilde{\bm f}_{\alpha
},\widetilde{\bm \gamma }_{\alpha }\right\} $ , and the source term
for inelastic collisions $\widetilde{\ell }$. All of the time
dependence is given by the evolution operator $\mathcal{U}_{\alpha
\beta }\left( t,T\right) $ which is that for the $N$ particle motion
in phase space, but compensated for all homogeneous dynamics.

\subsection{Green-Kubo Form to Order $k^{2}$}

Retaining only contributions up through order $k^{2}$ in Eq.\
(\ref{5.6f}) gives
\begin{eqnarray}
\mathcal{K}(n,T;{\bm k},t) & = & \mathcal{K}^{hyd} \left( n,T;{\bm
0}\right) -ik\left[ \widehat{\bm k} \cdot \overline{\bm D}(n,T;{\bm
0},0)+ \overline{Z}(n,T) \right] \nonumber  \\
&&+k^{2}\left[ \Lambda (n,T)+\overline{Y}(n,T) \right] , \label{5.7}
\end{eqnarray}
where the meaning of the different terms will be discussed next. The
first order in $k$ terms on the right hand side of this equation
provide the parameters for Euler order hydrodynamics. At this order,
the susceptibilities (pressure and pressure derivatives) are defined
in terms of the time independent correlation function $\overline{\bm
D}_{\alpha \beta }(n,T;{\bm 0},0)$, while the transport coefficient
$\zeta ^{U}$ is given by the Green-Kubo expression
\begin{equation}
\overline{Z}_{\alpha \beta}(n,T)= \delta_{\alpha 2} \delta_{\beta 3}
T \zeta^{U}(n,T), \label{5.7p}
\end{equation}
\begin{equation}
T\zeta ^{U}(n,T)=-\widehat{\bm k} \cdot \left[ \overline{\bm
S}_{23}^{(1)} (n,T;0)-\lim \int_{0}^{t}dt^{\prime}\overline{\bm
N}_{23}(n,T;{\bm 0,}t^{\prime })\right]. \label{5.7a}
\end{equation}
The above identification has been made by comparison of the
expression obtained here with the phenomenological transport matrix
in Eqs.\ (\ref{2.14})-(\ref{2.15}). Here and below, the notation for
Taylor series expansion of any function $X({\bm k})$  is
\begin{equation}
X({\bm k})=X({\bm 0})+i {\bm k} \cdot {\bm X}^{(1)}- {\bm k} {\bm k}
: {\sf X}^{(2)} + \ldots.   \label{5.7aa}
\end{equation}

At order $k^{2}$, the Navier-Stokes transport coefficients in Eq.
(\ref{5.7}) are of two types. The first type are those obtained from
$\Lambda (n,T)$, and represent the dissipative contributions to the
fluxes. They correspond  to the shear and bulk viscosities, thermal
conductivity, and $\mu $ coefficient in Eq.\ (\ref{2.6}). In the
Green-Kubo form they are determined by
\begin{equation}
\Lambda _{\alpha \beta }(n,T)=\widehat{\bm k} \widehat{\bm k} :
\left[ \overline{\sf D}_{\alpha \beta }^{(1)}(n,T;0)-\lim
\int_{0}^{t}dt^{\prime }\, \overline{\sf G}_{\alpha \beta
}(n,T;t^{\prime }) \right] ,  \label{5.7b}
\end{equation}
with
\begin{equation}
\overline{\sf G}(n,T;t) = \overline{\sf F}(n,T;{\bm
0},t)-\overline{\bm  D}(n,T;{\bm 0},0)\overline{\bm E} (n,T;{\bm 0},
t). \label{5.7bb}
\end{equation}
The transport coefficients of the second kind are those following
from $\overline{Y}(n,T)$ in Eq.\ (\ref{5.7}) and represent the
second order gradient contributions to the cooling rate, i.e. the
coefficients $\zeta ^{T}$ and $\zeta ^{n}$ in Eq.\ (\ref{2.4}),
\begin{equation}
\overline{Y}_{\alpha \beta }(n,T)=-\delta _{\alpha 2}\widehat{\bm
k}\widehat{\bm k} : \left[ \overline{\sf S}_{ 2 \beta
}^{(2)}(n,T;0)-\lim \int_{0}^{t}dt^{\prime }\, \overline{\sf
H}_{2\beta }(n,T;{\bm 0},t^{\prime })\right],  \label{5.7c}
\end{equation}
\begin{equation}
\overline{\sf H}_{2\beta }(n,T;{\bm 0},t )=\overline{\sf N}_{2\beta
}^{(1)}(n,T;t)+ T\zeta ^{U}\left( n,T\right) \widehat{\bm
k}\overline{\bm E}_{3\beta}  \left(n,T;{\bm 0},t \right).
\end{equation}
The superscripts $\left( 1\right) $ and $\left( 2\right) $ denote
coefficients in the expansion of correlation functions in powers of
$ik$, as indicated in Eq.\ (\ref{5.7aa}).

\subsection{Intermediate Helfand Form to Order $k^{2}$}

The intermediate Helfand form to order $k^{2}$ follows from direct
expansion of Eq.\ (\ref{5.5c}). The structure is the same as in Eq.\
(\ref{5.7}) as well as the contribution from $\overline{\bm
D}_{\alpha \beta }(n,T; {\bm 0},t)=\overline{\bm D}_{\alpha \beta
}(n,T; {\bm 0},0)$ . On the other hand, the transport coefficients
are now given by
\begin{equation}
T\zeta ^{U}(n,T)=-\lim \widehat{\bm k} \cdot \overline{\bm
S}_{23}^{(1)}(n,T;t), \label{5.8b}
\end{equation}
\begin{equation}
\Lambda (n,T)= \lim \widehat{\bm k} \widehat{\bm k} : \left[
\overline{\sf D}^{(1)}(n,T;t)-\overline{\bm D}(n,T;{\bm
0},0)\overline{\bm C}^{(1)}(n,T;t)\right] , \label{5.8c}
\end{equation}
\begin{equation}
\overline{Y}_{\alpha \beta}(n,T)=-  \delta_{\alpha 2} \lim  \left[
\widehat{\bm k} \widehat{\bm k}: \overline{\sf S} _{2\beta
}^{(2)}(n,T;t)+T\zeta ^{U}\left( n,T\right) \widehat{\bm k} \cdot
\overline{\bm C}_{3\beta }^{(1)}(n,T;t)\right].   \label{5.8d}
\end{equation}
The equivalence of these results with the Green-Kubo forms given in
the previous subsection can be seen by noting that the conservation
laws of Eqs.\ (\ref{5.6})-(\ref{5.6a}) to first order in $k$ give
\begin{equation}
\overline{\bm E}(n,T;{\bm 0,}t)=- \partial _{t}\overline{\bm
C}^{(1)}(n,T;t), \quad \overline{\sf F}(n,T;{\bm 0},t)=-
\partial _{t}\overline{\sf D}^{(1)}(n,T;t),
\label{5.9}
\end{equation}
\begin{equation}
\overline{\bm N}(n,T;{\bm 0},t)=-\partial _{t}\overline{\bm
S}^{(1)}(n,T;t),\quad \overline{\sf N}^{(1)}(n,T;t)=-\partial
_{t}\overline{\sf S}^{(2)}(n,T;t). \label{5.9a}
\end{equation}
These allows the time integrals in the Green-Kubo expressions to be
performed, giving directly Eqs.\ (\ref{5.8b})-(\ref{5.8d}).

\subsection{Einstein-Helfand Form to Order k$^{2}$}

Finally, the Einstein-Helfand form to order $k^{2}$ follows from
direct expansion of Eq.\, (\ref{6.10a}),
\begin{equation}
\mathcal{K}(n,T;{\bm k},t) =  \mathcal{K} \left( n,T;{\bm 0}\right)
+i {\bm k} \cdot \bm{\mathcal{K}}^{(1)}\left( n,T;t\right) - {\bm k}
{\bm k}: {\sf K}^{(2)}\left( n, T;t\right) , \label{5.9b}
\end{equation}
with
\begin{equation}
\bm{\mathcal{K}}^{(1)}\left( n,T;t\right) =-\left\{ \left[
\partial_{t}-\zeta _{0}\left( n,T\right) T\frac{\partial}{\partial T}+\mathcal{K}^{hyd}\left( n,T
;{\bm 0}\right) \right] \widetilde{\bm C}^{(1)}\left( n,T;t\right)
\right\} \widetilde{C}^{-1}\left( n,T;{\bm 0};t\right) \label{5.9c}
\end{equation}
and
\begin{eqnarray}
{\sf K}^{(2)}\left( n,T;t\right) &=&-\left\{ \left[
\partial_{t}-\zeta _{0}\left( n,T\right) T \frac{\partial}{\partial T}+\mathcal{K}^{hyd}\left(
n,T;{\bm 0}\right) \right] \widetilde{\sf C}^{(2)}(n,T;t)\right.  \nonumber \\
&&\left. +\bm{\mathcal{K}}^{(1)}\left( n,T;t\right) \widetilde{\bm
C} ^{(1)}(n,T;t)\right\} \widetilde{C}^{-1}\left( n,T,{\bm
0};t\right) . \label{5.9d}
\end{eqnarray}
This form does not separate explicitly the contributions leading to
the transport coefficients at both Euler and Navier-Stokes orders
for the cooling rate. In the previous two representations, this was
possible because the microscopic phase function $\widetilde{\ell
}(\Gamma ;n,T;{\bm k})$ due to energy loss in the inelastic
collisions appears explicitly.

\subsection{Dynamics and Projected Fluxes}

The Green-Kubo expressions involve the long time limit of time integrals
over correlation functions. This presumes the correlation functions decay
sufficiently fast for the integrals to exist. This decay time sets the time
scale after which the hydrodynamic description can apply. If these integrals
converge on that time scale then the Helfand formulas also reach \ their
limiting plateau values on the same time scale.

To explore this time dependence further, consider the correlation
function characterizing the transport coefficients $\Lambda _{\alpha
\beta }(n,T)$ associated with the heat and momentum fluxes (see Eq.
(\ref{5.7b})),
\begin{eqnarray}
\overline{\sf G}(n,T;t) &= &\overline{\sf F} (n,T;0,t)-\overline{\bm
D}(n,T;{\bm 0},0)
\overline{\bm E}(n,T;{\bm 0},t) \nonumber \\
&=&V^{-1} \int d\Gamma\, \widetilde{\bm f}(\Gamma
;n,T;{\bm 0})\left[ \widetilde{\bm \gamma }(\Gamma ;n,T;{\bm 0},t)\right. \nonumber \\
&&\left. -\widetilde{\psi}(\Gamma ;n,T;{\bm 0})V^{-1}\int d\Gamma\,
\widetilde{a}(\Gamma ;n,T;{\bm 0})\widetilde{\bm \gamma } (\Gamma ;n,T;{\bm 0},t)\right] \nonumber \\
& =& V^{-1}\int d\Gamma \widetilde{\bm f}(\Gamma ;n,T;{\bm 0}
)\left( 1-P\right) \widetilde{\bm \gamma }(\Gamma ;n,T;{\bm 0} ,t).
\label{5.10}
\end{eqnarray}
Then both contributions to $\overline{\sf G}$ combine to form the
projected part of the fluxes $\left( 1-P\right) \widetilde{\bm
\gamma }$, where $P$ is the projection onto the set $\left\{ \Psi
_{\alpha }(\Gamma ;n,T)\right\} $,
\begin{eqnarray}
PX(\Gamma) & =  & V^{-1} \Psi (\Gamma ;n,T)\int d\Gamma a(\Gamma
;n,T;{\bm
0})X(\Gamma)  \nonumber \\
& \equiv  & V^{-1} \sum_{\alpha} \Psi_{\alpha} (\Gamma ;n,T) \int
d\Gamma\, a_{\alpha} (\Gamma ;n,T;{\bm 0})X(\Gamma) . \label{5.11}
\end{eqnarray}
Thus $\left( 1-P\right) $ is a projection orthogonal to the
invariants, and $ \left( 1-P\right) \mathcal{U}\left( t,T\right)
P=0$, so
\begin{equation}
\left( 1-P\right) \mathcal{U}\left( t,T\right) \widetilde{\bm \gamma
}(\Gamma ;n,T)=\left( 1-P\right) \mathcal{U} \left( t,T\right)
\left( 1-P\right) \widetilde{\bm \gamma }(\Gamma ;n,T).
\label{5.11a}
\end{equation}
The time correlation function (\ref{5.10}) then can be rewritten as
\begin{eqnarray}
\overline{\sf G}_{\alpha \beta}(n,T;t)& = &V^{-1}\int d\Gamma\, {\bm
\Phi}_{\alpha}(\Gamma ;n,T) \left[ \mathcal{U}\left( t,T\right)
{\bm \Upsilon} (\Gamma ;n,T) \right]_{\beta} \nonumber \\
& = &  V^{-1}\sum_{\lambda} \int d\Gamma\, {\bm
\Phi}_{\alpha}(\Gamma ;n,T)  \mathcal{U}_{\beta \lambda}\left(
t,T\right) {\bm \Upsilon}_{\lambda} (\Gamma ;n,T), \label{5.11b}
\end{eqnarray}
where ${\bm \Phi}_{\alpha }$ and ${\bm \Upsilon} _{\alpha }$ are the
orthogonal fluxes
\begin{equation}
{\bm \Phi}_{\alpha }(\Gamma ;n,T)=\left( 1-P^{\dagger }\right)
\widetilde{\bm f} _{\alpha }(\Gamma ;n,T;{\bm 0}),\quad {\bm
\Upsilon} _{\alpha }(\Gamma ;n,T)=\left( 1-P\right) \widetilde{\bm
\gamma }_{\alpha }(\Gamma ;n,T), \label{5.11c}
\end{equation}
with the adjoint projection operator $P^{\dagger}$  given by Eq.\
(\ref{5.4a}).

Hence, there is no constant component of the correlation function
due to the invariants. Such a time independent part would not lead
to a convergent limit for the time integral, as required for the
transport coefficients. The property expressed by Eqs.\
(\ref{5.11c}) is similar to the presence of the ``subtracted
fluxes'' in the Green-Kubo expressions for the transport
coefficients of molecular fluids, with the subtracted fluxes being
orthogonal to the global invariants of the dynamics.

\section{Euler Order Parameters}
\label{sec6}

At Euler order, the unknown parameters of the phenomenological
hydrodynamics of Section \ref{sec2} are the cooling rate $\zeta
_{0}(n,T),$ the pressure $p(n,T)$, and the transport coefficient
$\zeta ^{U}$ associated with the expansion of the cooling rate to
first order in the gradients. The cooling rate has been already
defined by Eq.\ (\ref{4.3d}) above. The pressure and $\zeta ^{U}$
can also be given explicit definitions in terms of the correlation
functions from the coefficient of $k$ in Eq.\ (\ref{5.7}). Since
$\zeta ^{U}$ was previously identified as given given by Eq.\
(\ref{5.7a}), comparison of the remaining Euler coefficients in
Eqs.\ (\ref{2.14}) and (\ref{5.7}) gives
\begin{equation}
\left(
\begin{array}{cc}
\mathcal{K}^{hyd,(a)}_{1} & 0 \\
0 & 0
\end{array}
\right) \equiv \widehat{\bm k} \cdot \overline{\bm D} (n,T;{\bm
0},0), \label{5.12}
\end{equation}
\begin{equation}
\mathcal{K}^{hyd,(a)}_{1}= \left(
\begin{array}{ccc}
0 & 0 & n \\
0 & 0 & \frac{h-e_{0,n}n}{e_{0,T}}  \\
\frac{p_{n}}{nm} & \frac{p_{T}}{nm} & 0
\end{array}
\right). \label{5.12r}
\end{equation}
Using the definition in Eq.\ (\ref{5.5e}) and taking into account
once again Eq.\ (\ref{6.01}), it follows that
\begin{equation}
{\bm k} \cdot \overline{\bm D}_{\alpha \beta }(n,T;{\bm 0},0)
=V^{-1}\int d\Gamma \, \widehat{\bm k} \cdot {\bm f}_{\alpha
}(\Gamma ;n,T;{\bm 0}) \left[ \frac{\partial \rho _{h}\left(
\Gamma ;n,T,{\bm U}\right) }{\partial y_{\beta }}\right]_{{\bm
U}=0}.\label{5.12a}
\end{equation}
It is shown in Appendix \ref{ap5} that all the matrix elements of
(\ref{5.12r}) follow from (\ref{5.12a}) if the pressure is
identified as
\begin{equation}
p(n,T)\equiv  (Vd)^{-1} \int d\Gamma \rho _{h}\left(\Gamma;
n,T\right) \text{tr } {\sf H}(\Gamma ),  \label{5.12b}
\end{equation}
where $\text{tr } {\sf H} (\Gamma)$ is the volume integrated trace
of the microscopic momentum flux. Its detailed form is given by Eq.
(\ref{e.6a}) of Appendix \ref{ap5}. This is the second non-trivial
result of the linear response analysis here, providing the analogue
of the hydrostatic pressure for a granular fluid. It is possible to
show that Eq.\ (\ref{5.12b}) leads to $p(n,T)= nT$ in the low
density limit, but at finite density the dependence on temperature
and density of the pressure is determined by details of the HCS
distribution, rather than the Gibbs distribution. This is in
contrast to the appearance of $e_{0}\left( n,T\right) $ in Eq.\
(\ref{5.12}) which is a choice made in the definition of the
temperature. In general, there is no relationship of $p(n,T)$ to
$e_{0}\left( n,T\right) $ via thermodynamics, as for a normal fluid.

The transport coefficient $\zeta ^{U}$ represents dissipation due
to inelastic collisions proportional to $\nabla \cdot {\bm U}$. It
has no analogue for normal fluids, where the Euler hydrodynamics
is referred to as ``perfect fluid'' equations, since there is no
dissipation in that case. The simplest representation of
$\zeta^{U}$ is the intermediate Helfand form, Eq.\ (\ref{5.8b}).
More explicitly, it is shown in Appendix \ref{ap5} that it can be
expressed as
\begin{equation}
\zeta ^{U}(n,T)=\lim (VTe_{0,T}d)^{-1} \int d\Gamma\,  W\left(
\Gamma ;n,T\right)
e^{-\overline{\mathcal{L}}_{T}t}\mathcal{M}_{\zeta^{U}}\left(
\Gamma ;n,T\right),   \label{5.13}
\end{equation}%
with the source term $W\left( \Gamma ;n,T\right) $ defined by
\begin{equation}
W\left( \Gamma ;n,T\right)  \equiv - LE(\Gamma) - N \left[
\frac{\partial \left( e_{0,T}T\zeta _{0}\right) }{\partial n}
\right]_{e_{0}}  - E(\Gamma) \left[ \frac{\partial \left(
e_{0,T}T\zeta _{0}\right) }{\partial e_{0}} \right]_{n}.
\label{5.14}
\end{equation}
The phase function $\mathcal{M}_{\zeta ^{U}}\left( \Gamma
;n,T\right) $ is the conjugate momentum
\begin{equation}
\mathcal{M}_{\zeta ^{U}}\left( \Gamma ;n,T\right)  \equiv \int d{\bm
r}\, {\bm r} \cdot \left[ \frac{\delta \rho _{\ell h}}{\delta {\bm
U}({\bm r})} \right]_{\{y_{\beta}\}=\{n,T,{\bm 0}\}}
=-\sum_{s=1}^{N}{\bm q}_{s}\cdot \frac{\partial \rho_{h}
(\Gamma;n,T)}{\partial {\bm v}_{s}}. \label{5.15}
\end{equation}
The second equality makes use of the local HCS distribution  form
for the velocity dependence at uniform density and temperature,
\begin{equation}
\left[ \rho _{lh}\left( \Gamma | \left\{ y_{\beta }\right\} \right)
\right]_{\{y_{\beta}\}=\{n,T,{\bm U}({\bm r})\}} =\rho _{h}\left[
\left\{ {\bm q}_{r},{\bm v}_{r}- {\bm U}({\bm q}_{r})\right\};n,T
\right) . \label{5.16}
\end{equation}

The corresponding Green-Kubo form follows from a similar analysis of (\ref%
{5.7a}), or by direct integration of Eq.\ (\ref{5.13}),
\begin{eqnarray}
\zeta ^{U}(n,T) &=&\lim (VTe_{0,T}d )^{-1} \int d\Gamma\, W \left(
\Gamma;n,T\right) \mathcal{M}_{\zeta ^{U}}\left( \Gamma ;n,T\right)   \nonumber \\
&&-\lim  (VTe_{0,T}d)^{-1} \int_{0}^{t}dt^{\prime }\,  \int
d\Gamma\, W\left( \Gamma ;n,T\right)
e^{-\overline{\mathcal{L}}_{T}t^{\prime }}
\overline{\mathcal{L}}_{T}\mathcal{M}_{\zeta ^{U}}\left( \Gamma
;n,T\right). \label{5.17}
\end{eqnarray}
This completes the identification of the Euler order parameters of
the linearized hydrodynamic equations. Namely, exact expressions for
the pressure and $\zeta ^{U}$ in terms of correlation functions for
the reference HCS have been derived.

\section{Navier-Stokes Transport Coefficients}
\label{s7} The six transport coefficients at order $k^{2}$ can be
easily identified  in terms of elements of the  correlation
functions matrices  $\Lambda $ and $\overline{Y}$ introduced in Eq.\
(\ref{5.7}). The twelve intermediate Helfand and Green-Kubo forms
are given in Appendix \ref{ap5}. Only the shear viscosity is
discussed here in some detail. Consider first its intermediate
Helfand form. The analysis parallels closely that of $\zeta ^{U}$ in
Appendix \ref{ap5}, with the result
\begin{equation}
\eta =- \lim V^{-1}\int d\Gamma {\sf H}_{xy}(\Gamma
)e^{-\overline{\mathcal{L}}_{T}t} \mathcal{M}_{\eta }\left( \Gamma
;n,T\right) .  \label{5.18}
\end{equation}
Here, ${\sf H}_{ij}(\Gamma )$ is the volume integrated momentum flux
of Eq. (\ref{e.6a}) in Appendix \ref{ap6}, and $\mathcal{M}_{\eta }$
is the moment defined by
\begin{equation}
\mathcal{M}_{\eta }(\Gamma;n,T)= \int d{\bm r}x \left[
\frac{\partial \rho_{lh}}{\partial U_{y}({\bm r})}
\right]_{\{y_{\beta}\}=\{n,T,{\bm 0}\}}=-
\sum_{r=1}^{N}q_{rx}\frac{\partial}{\partial v_{ry}} \rho _{h}
(\Gamma;n,T). \label{5.18a}
\end{equation}
The $xy$ components occur here since the $x$ axis has been taken
along $\widehat{\bm k}$ and the $y$ axis along the transverse
direction $\widehat{\bm{e}}_{1}$, to simplify the notation. The
corresponding Green-Kubo form is
\begin{eqnarray}
\eta  &=&-\lim V^{-1}\int d\Gamma {\sf H}_{xy}(\Gamma
)\mathcal{M}_{\eta }\left(
\Gamma ;n,T\right)   \nonumber \\
&&+\lim \int_{0}^{t}dt^{\prime }V^{-1}\int d\Gamma {\sf
H}_{xy}(\Gamma )e^{- \overline{\mathcal{L}}_{T}t^{\prime
}}\overline{\mathcal{L}}_{T}\mathcal{M} _{\eta }\left( \Gamma
;n,T\right) .  \label{5.19}
\end{eqnarray}
In this case, the projection operators in Eq.\ (\ref{5.11b}) can
be omitted since their contributions vanish from symmetry, and the
dynamics is orthogonal to the invariants without such terms.

The transport coefficient $\zeta ^{U}$ vanishes for normal fluids,
but the shear viscosity remains finite, as it is well known. It is
instructive at this point to compare and contrast the results
(\ref{5.18}) and (\ref{5.19}) for normal and granular fluids.
Suppose from the outset a local equilibrium canonical ensemble
corresponding to the equilibrium $\rho _{c}(\Gamma)$, had been
used to generate the initial perturbations. Then, assuming
non-singular conservative forces,
\begin{equation}
\mathcal{M}_{\eta }\left( \Gamma ;n,T\right) = -
\sum_{r=1}^{N}q_{rx} \frac{\partial}{\partial v_{rx}}
\rho_{c}(\Gamma)= mT^{-1} \rho_{c}(\Gamma) \sum_{r=1}^{N}q_{rx}
v_{ry}= T^{-1} \rho_{c}(\Gamma) {\sf M}_{xy}, \label{5.20}
\end{equation}
with
\begin{equation}
{\sf M}_{xy} \equiv \sum_{r=1}^{N} q_{r,x} v_{r,y}. \label{5.20a}
\end{equation}
Then,
\begin{equation}
\overline{\mathcal{L}}_{T}\mathcal{M}_{\eta }\left( \Gamma
;n,T\right) =  T^{-1} \rho _{c}(\Gamma)  L {\sf M}_{xy} = T^{-1}
\rho _{c}(\Gamma) {\sf H}_{xy} (\Gamma). \label{5.21}
\end{equation}
Therefore, the intermediate Helfand and Green-Kubo expressions for a
normal fluid then become
\begin{equation}
\eta =-\lim (VT)^{-1}\left\langle {\sf H}_{xy}{\sf
M}_{xy}(-t)\right\rangle _{c}, \label{5.22}
\end{equation}
and
\begin{equation}
\eta =\lim  (VT)^{-1}\int_{0}^{t}dt^{\prime }\left\langle {\sf
H}_{xy}{\sf H}_{xy}(-t^{\prime })\right\rangle _{c}, \label{5.23}
\end{equation}
respectively. The brackets denote an equilibrium canonical ensemble
average, and the dynamics is that of the Liouville operator
$L=\overline{L}$. These are the familiar results that have been
studied and applied for more than forty years.

To make the comparison between the elastic and inelastic cases,
consider first the intermediate Helfand forms (\ref{5.18}) and
(\ref{5.22}). The similarity between the structure for the normal
and granular fluid results is striking, but the substantive changes
are significant. For the granular fluid, the equilibrium ensemble
has been replaced by the HCS ensemble. In addition, the Liouville
operator has been replaced by that including the nonconservative
force and $\overline{L}\neq L$. Finally, the generator for the
dynamics includes the effect of temperature cooling in the reference
HCS, $\overline{L} \rightarrow \overline{\mathcal{L}}_{T}=
\overline{L}-\zeta _{0}T\partial / \partial T$. These differences
manifest themselves in the Green-Kubo expressions in analogous ways.
The inclusion of nonconservative forces implies a time independent
contribution, the first term of Eq.\ (\ref{5.19}), which vanishes in
the elastic limit of Eq.\ (\ref{5.23}). Also, due to the change in
the ensemble, the two fluxes of the time correlation function differ
for a granular fluid, while both are momentum fluxes for a normal
fluid. Still the structure is such that these fluxes are orthogonal
to the invariants of the dynamics in both cases, so that the time
integrals can be expected to converge.

\section{Dimensionless Forms and Scaling Limit}

\label{s8} The analysis presented up to this point is quite general,
and the only restriction placed on the nature of the microdynamics
is that it be Markovian and the trajectories be invertible. These
restrictions are satisfied by most models used to describe the
interaction between granular particles. Examples are the Hertzian
contact force model \cite{walton}, the linear spring-dashpot model
\cite{walton}, and the system of inelastic hard spheres  with
impact-velocity dependent coefficient of restitution
\cite{poschel96}. The definition of these models and the associated
generators of phase space dynamics are shortly reviewed in Appendix
\ref{ap2}.

In general, there are two energy scales in the problem being
addressed here. One is the total energy per particle or,
equivalently, the cooling temperature $T_{h}\left( t\right) $. The
other energy scale is determined by a property of the specific
collision model, called $\epsilon $ in the following. For the
Hertzian spring case, it is the average compression energy of the
spring. For hard spheres, it is fixed by some characteristic
relative velocity in the dependence of the restitution coefficient
on the relative velocity of the colliding pair. It is useful to
reconsider the Liouville equation in a dimensionless form that
identifies these two different scales. In the limit that their ratio
$ \epsilon /T_{h}(t)$ is small, a special scaling form of the
results above is obtained. This limit is exact for hard spheres with
constant restitution coefficient ($\epsilon=0$), and makes precise
the conditions under which that idealized model may be approximately
valid for many states of interest.

Consider the Liouville equation (\ref{4.5a}),  and introduce the
dimensionless variables
\begin{equation}
{\bm q}_{r}^{\ast} =\frac{{\bm q}_{r}}{l }, \quad {\bm
v}^{\ast}_{r}=\frac{{\bm v}_{r}}{v_{0}(T)}, \quad
\epsilon^{\ast}=\frac{\epsilon}{m v_{0}^{2}(T)}, \label{7.4a}
\end{equation}
\begin{equation}
s=s(t,T), \label{7.4}
\end{equation}
where
\begin{equation}
v_{0}(T) \equiv \left( \frac{2T}{m} \right)^{1/2}
\end{equation}
is a thermal velocity, $l$ is the mean free path, and the function
$s(t,T)$ verifies the partial differential equation
\begin{equation}
\left( \frac{\partial s}{\partial t} \right)_{T} - \zeta_{0}(n,T) T
\left( \frac{\partial s}{\partial T} \right)_{t}
=\frac{v_{0}(T)}{l}, \label{7.4b}
\end{equation}
with the boundary condition $s(0,T)=0$. The corresponding
dimensionless distribution function is
\begin{equation}
\rho ^{\ast }\left( \Gamma ^{\ast };\epsilon ^{\ast },s\right) =[l
v_{0}\left( T\right)])^{Nd}\rho (\Gamma ;n,T,t), \quad \Gamma ^{\ast
} \equiv  \{ {\bm q}_{r}^{\ast }, {\bm v}_{r}^{\ast }\}.\label{7.5}
\end{equation}
The dependence on the (dimensionless) density has been omitted on
the left hand side to simplify the notation. In these variables, the
dimensionless Liouville equation takes the form
\begin{equation}
\partial _{s}\rho ^{\ast } \left( \Gamma ^{\ast };\epsilon ^{\ast },s\right)
+\overline{\mathcal{L}}^{\ast }\left( \Gamma^{\ast}; \epsilon ^{\ast
}\right) \rho ^{\ast }\left( \Gamma ^{\ast };\epsilon ^{\ast
},s\right)=0,  \label{7.7}
\end{equation}
where
\begin{equation}
\overline{\mathcal{L}}^{\ast }\left( \Gamma^{*}; \epsilon ^{\ast
}\right) \rho ^{\ast }=\zeta _{0}^{\ast }\left( \epsilon ^{\ast
}\right) \epsilon ^{\ast }\frac{\partial \rho ^{\ast }}{\partial
\epsilon ^{\ast }}+\frac{\zeta _{0}^{\ast }(\epsilon^{\ast})}{2}
\sum_{r=1}^{N} \frac{\partial}{\partial {\bm v}_{r}}\cdot \left(
{\bm v}_{r}^{\ast }\rho ^{\ast } \right)+\overline{L}^{\ast }\left(
\Gamma^{*}; \epsilon ^{\ast }\right) \rho ^{\ast }, \label{7.8}
\end{equation}
with $\overline{L}^{\ast }(\Gamma^{\ast};\epsilon)=l
\overline{L}(\Gamma)/v_{0}(T)$ and $\zeta_{0}^{\ast}=l
\zeta_{0}(T)/v_{0}(T)$.

To interpret this result further, it is useful to consider the
distribution of the HCS,
$\rho_{h}^{\ast}(\Gamma^{*};\epsilon^{\ast})$, which is the steady
state solution of Eq.\ (\ref{7.7}), i.e.,
\begin{equation}
\overline{\mathcal{L}}^{\ast }\left(\Gamma^{\ast}; \epsilon ^{\ast
}\right) \rho _{h}^{\ast }\left( \Gamma ^{\ast };\epsilon ^{\ast
}\right) =0. \label{7.9}
\end{equation}
This solution is the dimensionless form of the universal function
$\rho _{h}\left( \Gamma ;n,T\right) $, where the dependence on $T$
has been separated into a part that simply scales the velocities,
and a part that adimensionalizes the collisional energy scale
$\epsilon $. This shows that in the appropriate variables, the
distribution function of the HCS is stationary and universal, even
when velocity scaling alone (see below) does not hold. For an
isolated system, $\epsilon ^{\ast }(t) \equiv \epsilon / 2 T_{h}(t)$
grows with increasing $t$ since the system temperature decreases.
For very large $\epsilon ^{\ast }(t)$, the collisions become elastic
and the system approaches a normal fluid. However, the alternative
view of (\ref{7.7}) is to specify the solution as a function of
$(\Gamma ^{\ast },\epsilon ^{\ast }) $ and then study its properties
as a special non-equilibrium steady state of granular fluids.

As just noted, for large $\epsilon ^{\ast }$ the collisions become
practically elastic. In the opposite limit, $\epsilon ^{\ast }<<1$,
the dependence on $\epsilon ^{\ast }$ of the distribution function
can be neglected,
\begin{equation}
\rho ^{\ast }\left( \Gamma ^{\ast };\epsilon ^{\ast },s\right)
\rightarrow \rho ^{\ast }\left( \Gamma ^{\ast };s\right)
\label{7.10}
\end{equation}
and the Liouville equation becomes independent of $\epsilon ^{\ast
}$
\begin{equation}
\partial _{s}\rho ^{\ast }(\Gamma^{\ast})+
\overline{\mathcal{L}}^{\ast }(\Gamma^{\ast})\rho ^{\ast
}(\Gamma^{\ast})=0, \label{7.11}
\end{equation}
\begin{equation}
\overline{\mathcal{L}}^{\ast }(\Gamma^{\ast})\rho ^{\ast
}(\Gamma^{\ast})=\frac{\zeta _{0}^{\ast }}{2} \sum_{r=1}^{N}
\frac{\partial}{\partial {\bm v}^{\ast}_{r}} \cdot \left[{\bm
v}^{\ast }_{r} \rho ^{\ast }(\Gamma^{\ast})
\right]+\overline{L}^{\ast }\rho ^{\ast }(\Gamma^{\ast}).
\label{7.11a}
\end{equation}
This is the limit in which all temperature dependence occurs through
velocity scaling alone, as there is no other significant energy
scale. It occurs for sufficiently hard interactions and/or
sufficiently large kinetic energy. Many simplifications then occur.
The HCS solution has the simple form $\rho _{h}^{\ast }\left( \Gamma
^{\ast }\right) $ and consequently $ \zeta _{0}^{\ast }$ is a pure
number. Equation (\ref{7.4b}) defining $s$  can be integrated in
this case to give
\begin{equation}
s=-\frac{2}{\zeta _{0}^{\ast }}\ln \left[ 1-\frac{\zeta _{0}^{\ast
}v_{0}(T)t}{ 2l }\right].  \label{7.11b}
\end{equation}
Similarly, the dimensionless form of the cooling equation for
$T_{h}(t)$, Eq.\  (\ref{4.3a}), can be integrated to get the
explicit dependence on $t$
\begin{equation}
\frac{T_{h}(t)}{T_{h}(0)}=\left\{ 1+\frac{\zeta _{0}^{\ast
}v_{0}[T_{h}(0)]t}{ 2l }\right\} ^{-2}.  \label{7.12}
\end{equation}%
Finally, when (\ref{7.11a}) is evaluated at $T_{h}(t)$ the
relationship of $s $ to $t$ becomes
\begin{equation}
s=\frac{2}{\zeta _{0}^{\ast }}\ln \left[ 1+\frac{v_{0}(0)}{2 l
}\zeta _{0}^{\ast }t\right] .  \label{7.13}
\end{equation}%
The conditions for which Eq.\ (\ref{7.10}) applies will be called
the ``scaling limit''.

The special collisional model of inelastic hard spheres with
constant restitution coefficient has no intrinsic collisional energy
scale, so $ \epsilon ^{\ast }=0$ and the scaling limit is exact. The
generators for this dynamics are indicated in Appendix \ref{ap2} and
can be understood as the singular limit of a soft, continuous
potential, like the Hertzian contact force model. The analysis of
the preceding sections is specialized to this case in the following
paper \cite{BDB06}, where it is shown that the above simplifications
admit a more detailed exposition of the formal expressions for the
transport coefficients. The rest of this section is a brief
translation of some of the main results here to their dimensionless,
scaling limit form.

The dimensionless hydrodynamic fields are defined by
\begin{equation}
\left\{ \delta y^{\ast}_{\alpha} \right\} \equiv  \left\{
\frac{\delta y_{\alpha}}{\overline{y}_{\alpha,h}} \right\} \equiv
\left\{ \frac{\delta n}{n_{h}}\, , \frac{\delta T}{T_{h}}\, ,
\frac{\delta{\bm U}}{v_{0}(T)} \right\}, \label{7.13aa}
\end{equation}
where the definition of the $\overline{y}_{\alpha,h}$'s follows from
the second identity. Then, the fundamental linear response equation,
Eq. (\ref{6.7}), in the dimensionless variables $\delta
\widetilde{y}_{\alpha }^{\ast }$ is
\begin{equation}
\delta \widetilde{y}^{\ast }({\bm k}^{\ast}, s)=\widetilde{C}^{\ast
}\left( {\bm k}^{\ast },s\right) \delta \widetilde{y}^{\ast }({\bm
k}^{\ast},0) \label{7.14},
\end{equation}
with
\begin{equation}
\widetilde{C} _{\alpha \beta }^{\ast } ({\bm k}^{\ast },s ) =
\frac{1}{ \overline{y}_{\alpha, h}\left[ T_{h}(t)\right]
}\widetilde{C}_{\alpha \beta }\left[ n_{h},T_{h}(t);{\bm k},t\right]
\overline{y}_{\beta, h}\left[ T_{h}\left( 0\right) \right].
\label{7.14a}
\end{equation}
It follows from Eq.\ (\ref{6.8}) that $\widetilde{C}_{\alpha \beta
}^{\ast }\left( {\bm k}^{\ast },s\right) $ obeys the equation
\begin{equation}
\left[ \partial _{s}+\mathcal{K}^{\ast }({\bm k}^{\ast },s)\right]
\widetilde{C}^{\ast }\left( {\bm k}^{\ast },s\right) =0, \quad
\widetilde{C}^{\ast }\left( {\bm k}^{\ast},0 \right) =I,
\label{7.15}
\end{equation}
the dimensionless transport matrix being
\begin{equation}
\mathcal{K}_{\alpha \beta }^{\ast }\left({\bm k}^{\ast },s\right) =
-\delta _{\alpha \beta}p_{\alpha }\zeta ^{\ast }_{0}+\frac{l
\overline{y}_{\beta, h }\left[ T(t) \right]}{ v_{0}\left[ T_{h}(t)
\right] \overline{y}_{\alpha,h  }\left[ T(t) \right]
}\mathcal{K}_{\alpha \beta }\left( n,T;{\bm k},t\right)  .
\label{7.16}
\end{equation}
Here $\mathcal{K}\left( n,T;{\bm k},t\right) $ is the transport
matrix analyzed in the previous sections. The additional
contributions to $\mathcal{K}^{\ast }({\bm k}^{\ast },s)$,
proportional to $\left\{ p_{\alpha } \right\} \equiv \left\{
0,1,\frac{1}{2},\ldots ,\frac{1}{2}\right\} $ arise from
differentiating the normalization constants with respect to $T$.
Because of the scaling limit, $\mathcal{K}_{\alpha \beta }^{\ast
}\left( {\bm k}^{\ast },s\right) $ is independent of $T(t)$ and the
hydrodynamic limit can be identified as
\begin{equation}
\mathcal{K}^{\ast hyd }({\bm k}^{\ast })=\mathcal{K}^{\ast }({\bm k}
^{\ast },\infty ).  \label{7.17}
\end{equation}
The phenomenological form of $\mathcal{K}^{\ast hyd }({\bm k}^{\ast
})$, corresponding to that of Sec.\  \ref{sec2} above, is given in
the following companion paper \cite{BDB06}.

The dimensionless forms for the response functions
$\widetilde{C}_{\alpha \beta }^{\ast }\left( {\bm k}^{\ast
},s\right) $ as phase space averages, follow from Eq.\ (\ref{6.8a}),
\begin{eqnarray}
\widetilde{C}_{\alpha \beta }^{\ast }\left( {\bm k}^{\ast },s\right)
&=&\left[ \frac{1}{V \overline{y}_{\alpha, h}\left( T\right) }\int
d\Gamma\, \widetilde{a}_{\alpha }\left( \Gamma ;n,T,{\bm k}\right)
e^{- t \overline{ \mathcal{L}}_{T}}\widetilde{\psi }_{\beta }(\Gamma
;n,T,-{\bm k}
) \right]_{n=n_{h},T=T_{h}(t)} \overline{y}_{\beta, h}\left[ T(0)\right] \nonumber \\
&= & \frac{1}{V y_{\alpha, h}[T_{h}(t)]} \int d \Gamma\,
\widetilde{a}_{\alpha} \left[ \Gamma;n_{h},T_{h}(t);{\bm k} \right]
e^{- t \overline{L}} \widetilde{\psi´}_{\beta} (\Gamma;
n_{h},T_{h}(0); -{\bm k}) \overline{y}_{\beta,h} \left[ T_{h}(0)
\right] \nonumber \\
&=& V^{\ast -1} \int d\Gamma^{\ast}\,  \widetilde{a}_{\alpha }^{\ast
}\left( \Gamma ^{\ast }; {\bm k}^{\ast }\right) e^{-s
\overline{\mathcal{L}}^{\ast }}\widetilde{ \psi }_{\beta }^{\ast
}(\Gamma ^{\ast };-{\bm k}^{\ast }), \label{7.18}
\end{eqnarray}
where $V^{\ast} = V/ l^{d}$. More details of this transformation are
given in Appendix B of ref. \cite{BDB06}. The generator for the
dynamics $\overline{\mathcal{L}}^{\ast }$ is given by Eq.\
(\ref{7.11}), and the dimensionless phase functions are
\begin{equation}
\widetilde{a}_{\alpha }^{\ast }\left( \Gamma ^{\ast };{\bm k}^{\ast
}\right) =\frac{\widetilde{a}_{\alpha }\left( \Gamma ;n,T,{\bm k}
\right) }{l^{d} \overline{y}_{\alpha, h}\left( T\right) },
\label{7.19a}
\end{equation}
\begin{equation}
\widetilde{\psi } _{\beta }^{\ast }(\Gamma ^{\ast };{\bm k}^{\ast
})=\left[ v_{0}(T) l \right] ^{N d}\widetilde{\psi }_{\beta }(\Gamma
;n,T;{\bm k})\overline{y}_{\beta, h}\left( T\right) .  \label{7.19}
\end{equation}
The hydrodynamic transport matrix $\mathcal{K}^{\ast hyd }({\bm
k}^{\ast }) $ is therefore given by
\begin{equation}
\mathcal{K}^{\ast hyd }({\bm k}^{\ast })=\mathcal{K}^{\ast hyd}({\bm
0} )-\lim \left\{ \left[ \partial _{s}+\mathcal{K}^{\ast hyd }({\bm
0})\right] \widetilde{C}^{\ast }\left( {\bm k}^{\ast },s\right)
\right\} \widetilde{C} ^{\ast -1}\left({\bm k}^{\ast },s\right) ,
\label{7.20}
\end{equation}
which is the dimensionless form of Eq.\ (\ref{6.10a}). In this way,
all relation to the cooling of the reference state through a
dependence on $T$ has been removed. However, the dynamics of
homogeneous perturbations of this state remains through
\begin{equation}
\widetilde{C}^{\ast }\left( {\bm 0},s\right) =e^{-\mathcal{K} ^{\ast
hyd }({\bm 0})s}.  \label{7.21}
\end{equation}

The dimensionless correlation functions defining the transport
coefficients are obtained in a similar way and have representations
analogous to (\ref {7.18}). An important difference is that the
generator is $\overline{ \mathcal{L}}^{\ast }-\mathcal{K}^{hyd\ast
}({\bm 0})$, indicating that the homogenous hydrodynamics is
compensated. These simplifications and further interpretation are
also deferred to the following paper.

\section{Discussion}

The objective of this work has been to translate the familiar
methods of linear response for normal fluids to the related, but
quite different case, of granular fluids. In both cases, the linear
response to perturbations of a homogeneous reference state is
described in terms of the fundamental tools of non-equilibrium
statistical mechanics. This microscopic formulation is then compared
with the corresponding description from phenomenological
hydrodynamics, and the unknown parameters of the latter are
identified in terms of associated response functions. The analysis
entails several steps, and at each stage there are technical and
conceptual differences encountered for granular fluids that have
been addressed in the preceding sections.

The first difference is the form of the reference homogeneous state.
For normal fluids, this is the equilibrium Gibbs state, while for
granular fluids the corresponding role is played by the HCS. In both
cases, the ideal global case considered  is thought to be
representative of more general cases, where these global states are
expected to represent the state locally. The Gibbs state is strictly
stationary and therefore constructed from the dynamical invariants.
The HCS has an inherent time dependence due to collisional energy
loss (``cooling''). Here, it has been given a stationary
representation by including the granular temperature as a dynamical
variable. In this form, the granular linear response problem becomes
similar to that for a normal fluid, namely the response to the
spatial perturbation of a homogeneous, stationary state. However,
the generators for that dynamics are now more complex, due both to
the non-conservative forces responsible for collisional energy loss
and a generator for changes in the granular temperature.

The response functions contain information about both hydrodynamics
and microscopic collective excitations. Therefore, the initial
perturbation will excite in general a wide range of dynamics and the
extraction of the hydrodynamic branch at long wavelengths, long
times, can be quite complex. However, if the perturbation is chosen
to excite only the hydrodynamic modes, this analysis becomes simpler
and more direct. Practically, this can be done only in the long
wavelength limit. For a normal fluid, these are the dynamical
invariants, since the hydrodynamic modes are those that vanish in
that limit. The associated perturbation is given by the
corresponding global conserved quantities (number, energy,
momentum). The granular fluid is more complex, since there is a
residual hydrodynamics even in the long wavelength limit. This is
due to the nonlinear temperature cooling that is linearized about a
reference cooling state. The hydrodynamic modes are therefore
identified from this non-zero long wavelength dynamics. The
microscopic perturbations corresponding to this dynamics are
described in Sec. \ref{sec3}. Furthermore, by extracting this
dynamics from the the microscopic evolution, these perturbations
become the invariants of the residual dynamics (see Eqs.
(\ref{4.13}) and (\ref{4.14})).

With this knowledge of the special long wavelength hydrodynamic
perturbations, spatial perturbations are constructed from their
corresponding local forms. In the normal fluid case, these are the
microscopic local densities of number, energy, and momentum, and are
generated by a local equilibrium ensemble. By analogy, the densities
of the invariants for the granular fluid are the appropriate
perturbations, and they are generated from a corresponding local
HCS. Since the HCS is not the Gibbs state, these densities are no
longer the local conserved densities for a normal fluid.

The remaining analysis is straightforward, extracting the Euler and
Navier-Stokes parameters as coefficients in a wavevector expansion,
using the conservation laws to assist in the ordering. There are two
different sets of conservation laws (balance equations) for the
granular fluid, in contrast to only one set for the normal fluid.
Consequently, expressions for transport coefficients are not simply
given in terms of autocorrelation functions of appropriate fluxes.
Instead, for the granular fluid there are two conjugate fluxes in
the Green-Kubo expressions.

The utility of these results rests on further studies of appropriate
ways to evaluate them. The status now is that exact expressions for
the quantities of interest (e.g., pressure and transport
coefficients) are given formally without any inherent uncontrolled
approximations (e.g., as in some chosen kinetic theory). This is a
more suitable point for the introduction of practical methods for
evaluation. In the study of normal fluids, a number of methods have
proved very useful. They include molecular dynamics simulations,
memory function models incorporating exact initial dynamics, and
linear kinetic theory. It is hoped that similar approaches will be
developed for the expressions provided here. To elaborate on this,
consider again the new Euler order transport coefficient given by
Eq.\ (\ref{5.13}), make the temperature scaling of the generator
explicit, and evaluate the entire expression at $T=T_{h}(t)$
\begin{eqnarray}
\zeta ^{U}\left[ n,T_{h}(t) \right] & = &- \lim \left[
(VTe_{0,T}d)^{-1} \int d\Gamma\, W\left( \Gamma ;n,T\right)
e^{-\overline{L}t}e^{t\zeta _{0}T \frac{\partial}{\partial T}}
\sum_{s=1}^{N} {\bm q}_{s} \cdot \frac{\partial \rho
\left(\Gamma;n,T \right)}{\partial {\bm v}_{s}} \right]_{T=T_{h}(t)}
\nonumber \\
& = & - \lim \left [V T_{h}(t)e_{0,T}\left[ T_{h}(t)\right]  d
\right]^{-1}\int d\Gamma\,  W\left[ \Gamma ;n,T_{h}(t)\right]
e^{-\overline{L}t}  \sum_{s=1}^{N} {\bm q}_{s} \cdot \frac{\partial
\rho \left[\Gamma;n,T(0) \right]}{\partial {\bm v}_{s}}.
\nonumber \\
\label{8.2}
\end{eqnarray}

The generator for the dynamics is now that for the trajectories
alone, so this is a form suitable for MD simulation. The simulation
method must be constructed in such a way as to represent the unknown
HCS appearing in (\ref {8.2}). Elsewhere, the evaluation by kinetic
theory \cite{BDB206} is considered, and  the original representation
given by Eq.\ (\ref{5.13}) is found to be more suitable.

It is worth recalling that liquid state transport for simple atomic
fluids remains a prototypical strongly coupled many-body problem,
with little progress beyond simulation of formal expressions such as
those given here. More should not be expected for ``complex''
granular fluids. The formal representations of transport
coefficients by methods of statistical mechanics provides a new
perspective on a difficult old problem. As for normal fluids,
significant further progress can be expected for the idealized model
of hard spheres. That is the subject of the following companion
paper.

\section{Acknowledgements}

The research of J.D. and A.B. was supported in part by the
Department of Energy Grant (DE-FG03-98DP00218). The research of
J.J.B. was partially supported by the Ministerio de Educaci\'{o}n y
Ciencia (Spain) through Grant No. BFM2005-01398 (partially financed
with FEDER funds). This research also was supported in part by the
National Science Foundation under Grant No. PHY99-0794 to the Kavli
Institute for Theoretical Physics, UC Santa Barbara. A.B. also
acknowledges a McGinty Dissertation Fellowship and IFT Fellowship
from the University of Florida.
\bigskip

\appendix

\section{Homogeneous State Dynamics}
\label{ap1} The dynamics associated with the homogeneous cooling
state of interest here is two fold. The first is the cooling of the
temperature, determined from the solution to Eq.\ (\ref{2.7aa}).
For a given initial condition $T$, the solution is denoted by
$T_{h}\left( t;n_{h},T\right) $. The density is a constant parameter
and sometimes it is left implicit in the notation of the text. The
second dynamics is the linear response to small homogeneous changes
in the initial conditions,
\begin{equation}
\delta T_{h}\left( t;n_{h},T\right) = \left( \frac{\partial
T_{h}\left( t;n_{h},T\right) }{\partial n_{h}} \right)_{T} \delta
n_{h}+ \left( \frac{\partial T_{h}\left( t;n_{h},T\right) }{\partial
T}\right)_{n_{h}} \delta T. \label{a.2}
\end{equation}
In this Appendix, it is shown how the response function for this
second type of dynamics is obtained from the linearized hydrodynamic
equations to give  Eq.\ (\ref{2.23a}).

A useful identity for any function of the temperature, $X(T)$ is
\begin{equation}
X[T(t_{2})]=\exp \left\{ -\left( t_{2}-t_{1}\right) \zeta_{0}
\left[ T\left( t_{1}\right) \right] T\left( t_{1}\right)
\frac{\partial }{\partial T\left( t_{1}\right) }
\right\}X[T(t_{1})]. \label{a.4}
\end{equation}
Here $T(t)$ is a solution to Eq.\ (\ref{2.7aa}). The identity can
be proved by performing a Taylor series of $X[T(t_{2})]$ in powers
of $\left( t_{2}-t_{1}\right) $ and using Eq.\ (\ref{2.7aa}) to
evaluate the time derivatives in terms of $T(t_{1})$ derivatives.
The above identity gives, in particular,
\begin{eqnarray}
T(0) &= & \exp \left\{ t\zeta_{0} \left [ T\left( t\right) \right]
T\left( t\right) \frac{
\partial }{\partial T\left( t\right) } \right\}  T(t),\nonumber \\
T(t) & =  & \exp \left\{-t\zeta_{0} \left[ T\left( 0\right) \right]
T\left( 0\right) \frac{\partial }{\partial T\left( 0\right) }
\right\}T(0), \label{a.5}
\end{eqnarray}
and
\begin{equation}
\left( \frac{\partial T(t)}{\partial
T(0)}\right)_{n_{h}}=\frac{\zeta_{0} \left[ T\left( t\right) \right]
T\left( t\right) }{\zeta_{0} \left[ T\left( 0\right) \right] T\left(
0\right)}\, .  \label{a.6}
\end{equation}

Consider some function $X[T(t;T),t]$ that depends on time through
$T(t;T)$ plus some residual time dependence. Use the second
equation of (\ref{a.5}) to write
\begin{equation}
X\left[ T(t;T),t\right] = \exp \left[ -t\zeta_{0} \left( T\right)
T \frac{\partial}{\partial T} \right] X(T,t) \label{a.7}
\end{equation}
and, consequently,
\begin{eqnarray}
\partial _{t}X \left[ T(t;T),t \right]  &=& \exp \left[ -t\zeta_{0} \left( T\right) T
\frac{\partial}{\partial T} \right] \left\{ \left[ \partial
_{t}-\zeta_{0} \left( T\right) T \frac{\partial}{\partial T} \right]
X(T,t)
\right\}  \nonumber \\
&=&\left\{ \left[ \partial _{t}-\zeta_{0} \left( T\right) T
\frac{\partial}{\partial T} \right] X(T,t)\right\} _{T=T(t;T)}\, .
\label{a.8}
\end{eqnarray}
The time dependence due to $T(t;T)$ can be replaced by treating $T$
as an independent variable, with the additional generator for its
dynamics $\zeta_{0} \left( T\right) T \partial / \partial T$. It is
then equivalent to determine $X(T,t)$, and evaluate it finally at
$T=T(t;T).$ In the case of Eq. (\ref{2.18}), this leads to
\begin{equation}
\left[ \partial _{t}-\zeta_{0} \left( T\right) T
\frac{\partial}{\partial T}+ \mathcal{K} ^{hyd}\left( n,T;{\bm
k}\right) \right] \widetilde{C}^{hyd}\left( n,T;{\bm k},t\right) =0,
\label{a.9}
\end{equation}
\begin{equation}
\widetilde{C}_{\alpha \beta }^{hyd}\left( n,T;{\bm k},0\right)
=\delta _{\alpha \beta },
\end{equation}
with the definition in Eq.\ (\ref{2.20}).  For  ${\bm k}={\bm 0}$,
use of Eqs.\ (\ref{2.14})-(\ref{2.15}) gives
\begin{equation}
\widetilde{C}_{\alpha \beta }^{hyd}\left( n,T;{\bm 0},t\right)
=\delta _{\alpha \beta },  \label{a.10}
\end{equation}
for $\alpha \neq 2$, while for $\alpha =2$ it is found:
\begin{equation}
\left[ \partial _{t}-\zeta_{0} \left( T\right) T
\frac{\partial}{\partial T}+\mathcal{K} _{22}^{hyd}\left( n,T;{\bm
0}\right) \right] \widetilde{C}_{2\beta }^{hyd}\left( n,T;{\bm
0},t\right) +\mathcal{K}_{21}^{hyd}\left( n,T; {\bm 0}\right)
\delta _{1\beta }=0,  \label{a.11}
\end{equation}
or, more explicitly,
\begin{equation}
\left[  \partial _{t}-\zeta_{0} T \frac{\partial}{\partial T}
+\left( \frac{\partial \left( \zeta_{0}  T\right) }{\partial T}
\right)_{n}\right] \widetilde{C}_{2\beta }^{hyd}\left( n,T;{\bm
0},t\right) +\left( \frac{\partial \left( \zeta_{0} T\right)
}{\partial n} \right)_{T} \delta _{1\beta }=0. \label{a.12}
\end{equation}
The solution of this equation is
\begin{equation}
\widetilde{C}_{2\beta }^{hyd}\left( n,T;{\bm 0},t\right) = \left(
\frac{\partial T }{\partial n} \right) _{T(-t;T)}\delta _{1\beta }+
\left( \frac{\partial T}{\partial T(-t;T)}\right) _{n}\delta
_{2\beta }, \label{a.13}
\end{equation}
as can be verified by direct substitution into Eq.\ (\ref{a.12})
and repeated use of Eq.\ (\ref{a.6}). This is the result
(\ref{2.23a}) of the text.

\section{Generators of Dynamics}
\label{ap2}

The interaction between the constituent particles of the dissipative
fluid enters the presentation  here via the Liouville operators that
generate the dynamics. The analysis of the text places few
restrictions on these generators and admits a large class of models
to represent real systems. For example, it is not necessary that
they be pairwise additive, although the examples of this appendix
all assume that case. There is a qualitative difference between the
generators for continuous or piecewise continuous forces, and those
for singular forces (e.g., hard spheres). Examples of each are given
here for illustration.

\subsection{Dissipative soft spheres}

The fluid is assumed to be comprised of mono-disperse spherical
particles with pairwise additive central interactions. The latter
implies that the forces are ``smooth'', without tangential
momentum transfer, and Newton's third law holds. The simplest
realistic model for the force ${\bm F}$ that particle $s$ exerts
on particle $r$ is the smooth, frictional contact model
\cite{walton, poschel96} given by
\begin{equation}
\mathbf{F}\left( {\bm q}_{rs},{\bm g}_{rs}\right) =\widehat{\bm
q}_{rs}\Theta \left( \sigma -q_{rs}\right) \left[ f\left( \sigma
-q_{rs}\right) -\gamma \left( \sigma -q_{rs}\right) \left( {\bm g}
_{rs}\cdot \widehat{\bm q}_{rs}\right) \right] .  \label{b.1}
\end{equation}
where $\Theta (x)$ is the Heaviside step function, ${\bm q}_{rs}
={\bm q}_{r}-{\bm q} _{s}$  is the relative coordinate, ${\bm
g}_{rs}={\bm v}_{r}-{\bm v}_{s}$ is the relative velocity of the two
particles, and $\widehat{\bm q}_{rs} \equiv {\bm q}_{rs}/q_{rs}$ is
the unit normal vector joining the centers of the two particles.
Moreover, $f(x)$ and $\gamma$ are a function and a constant,
respectively, to be described below. This is a piecewise continuous
force that vanishes for separations greater than $\sigma $, which
therefore can be thought of as the diameter of the particles. The
first term between the brackets describes  a \emph{conservative}
force representing the elastic repulsion due to the deformation of
real granular particles. If $f\left( x\right) $ is chosen to be
linear, the deformation is that of a spring. The amount of
deformation can be adjusted by the choice of the spring constant. A
second, more realistic, choice is the Hertzian contact model for
which $f\left( x\right) \propto x^{3/2}$.

The second term of (\ref{b.1}) is a \emph{nonconservative} force
representing the energy loss of the particle pair on collision. It
is proportional to the relative velocity of approach during the
collision, and the amount of energy loss is adjusted by the choice
of the friction constant $ \gamma $.

The Liouville operators $L$ and $\overline{L}$, defined in Eqs.
(\ref{3.2}) and (\ref{3.3}), for the dynamics of phase functions
and distributions for these models can be identified as
\begin{equation}
LX\left( \Gamma \right) \equiv \sum_{r=1}^{N} {\bm v}_{r}\cdot
\frac{\partial}{\partial {\bm q}_{r}} X\left( \Gamma \right)
+\frac{1}{m}\sum_{r=1}^{N} \sum_{s \neq r}^{N} {\bm F}\left( {\bm
q}_{rs},{\bm g}_{rs}\right) \cdot \frac{\partial}{\partial {\bm
v}_{r}} X\left( \Gamma \right)  \label{b.2}
\end{equation}
and
\begin{equation}
\overline{L}X\left( \Gamma \right) \equiv LX\left( \Gamma \right)
+\frac{1}{m} \sum_{r=1}^{N} \sum_{ r \neq s}^{N} X(\Gamma)
\frac{\partial}{\partial {\bm v}_{r}} \cdot {\bm F} \left( {\bm
q}_{rs},{\bm g}_{rs}\right). \label{b.3}
\end{equation}
It is readily verified that the total momentum is conserved, since
Newton's third law is satisfied, i.e.,  ${\bm F}\left({\bm
q}_{rs},{\bm g} _{rs}\right) =-{\bm F}\left( {\bm q}_{sr},{\bm
g}_{sr}\right) $. The total energy is
\begin{equation}
E(\Gamma)
=\sum_{r=1}^{N}\frac{1}{2}mv_{r}^{2}+\frac{1}{2}\sum_{r=1}^{N}
\sum_{s \neq r}^{N} V\left( q_{rs}\right) , \label{b.4}
\end{equation}
where the potential energy function $V(q)$ verifies
\begin{equation}
\frac{\partial V\left( q_{rs}\right) }{\partial q_{rs}}=- \Theta
\left( \sigma -q_{rs}\right) f\left( \sigma -q_{rs}\right).
\label{b.4b}
\end{equation}
The microscopic energy loss is easily computed by using Eq.\
(\ref{3.5}) with the result
\begin{equation}
LE (\Gamma) =-\frac{1}{2}\sum_{r=1}^{N} \sum_{s \neq r}^{N} \Theta
\left( \sigma -q_{rs}\right) \gamma \left( \sigma -q_{rs}\right)
\left( {\bm g}_{rs}\cdot \widehat{{\bm q}}_{rs}\right) ^{2},
\label{b.5}
\end{equation}
showing that it is associated with the nonconservative part of the
interactions as it should.

\subsection{Hard sphere dynamics}

For a given energy of activation, the contact forces considered
above may have small deformations, i.e. the region in which the
forces differs from zero verifies  $\left( \sigma -q_{rs}\right)
/\sigma \ll 1$. In that case, the conservative part of the force
approaches that of elastic hard spheres. The primary effect of the
nonconservative force is to decrease the magnitude of $ {\bm
g}_{rs}\cdot \widehat{\bm q}_{rs}$ after the collision. This can be
represented by the scattering law
\begin{equation}
{\bm g}^{\prime}_{rs}={\bm g}_{rs}-\left[ 1+\alpha \left(
g_{rs}\right) \right] \left( \widehat{\bm \sigma} \cdot {\bm
g}_{rs}\right) \widehat{\bm \sigma}, \label{b.7}
\end{equation}
where\ ${\bm g}^{\prime}_{rs}$ is the relative velocity after
collision and $\alpha \left( g_{rs}\right) $ is a coefficient of
restitution that depends on the relative velocity. The total
momentum of the pair is, by definition, unchanged in the
collision. The elastic limit corresponds to $\alpha \left(
g_{rs}\right) \rightarrow 1$. Subsequent to the change in relative
velocity for the pair $ (r,s)$, the free streaming of all
particles continues until another pair is at contact, and the
corresponding instantaneous change in their relative velocities is
performed. The collision rule is assumed to be invertible, i.e.,
$\alpha \left( g_{rs}\right) $ is specified so that the trajectory
can be reversed.

Since there is no longer a potential energy, the total energy for
the system is its kinetic energy, which changes on a pair collision
by
\begin{equation}
\Delta \left[ \frac{1}{2}m\left( v_{r}^{2}+v_{s}^{2}\right) \right]
=\frac{1}{4}m\left( {g}_{rs}^{\prime 2}-g_{rs}^{2}\right)
=-\frac{1}{4}m\left[ 1-\alpha ^{2}\left( g_{rs}\right) \right]
\left( \widehat{\bm \sigma} \cdot {\bm g}_{rs} \right) ^{2}.
\label{b.8}
\end{equation}
This is clearly the analogue of the terms on the right hand side of
Eq.\ (\ref {b.5}). In fact, the velocity dependence of the
coefficient of restitution can be modeled from a comparison of the
two equations.

There are two components of the generators $L$ and $\overline{L}$,
corresponding to each of the two steps of free streaming and
velocity changes at contact,
\begin{equation}
L=\sum_{r=1}^{N} {\bm v}_{r}\cdot \frac{\partial}{\partial {\bm
r}_{r}}+\frac{1 }{2}\sum_{r=1}^{N} \sum_{s \neq r}^{N}T(r,s),
\label{b.9}
\end{equation}
\begin{equation}
\overline{L}=\sum_{r=1}^{N}{\bm v}_{r}\cdot \frac{\partial}{\partial
{\bm r}_{r}}-\frac{1}{2}\sum_{r=1}^{N} \sum_{s \neq r}^{N}
\overline{T}(r,s). \label{b.10}
\end{equation}
The operators $T(r,s)$ and $\overline{T}(r,s)$ describe the binary
collision for a pair,
\begin{equation}
T(r,s)=\delta (q_{rs}-\sigma )\Theta (-{\bm g}_{rs}\cdot
\widehat{\bm q}_{rs})|{\bm g}_{rs}\cdot \widehat{\bm
q}_{rs}|(b_{rs}-1), \label{b.11}
\end{equation}
\begin{equation}
\overline{T}(r,s)=\delta (q_{rs}-\sigma )\left[ J\left( {\bm
v}_{r},{\bm v}_{s}\right) b_{rs}^{-1}-1 \right]\Theta (-{\bm
g}_{rs}\cdot \widehat{\bm q} _{rs})|{\bm g}_{rs}\cdot \widehat{\bm
q}_{rs}|.  \label{b.12}
\end{equation}
Here $b_{rs}$ is a substitution operator,
\begin{equation}
b_{rs}X({\bm g}_{rs})=X(b_{rs}{\bm g}_{rs})=X({\bm
g}_{rs}^{\prime}),\label{b.13}
\end{equation}
which changes the relative velocity ${\bm g}_{rs}$ into its
scattered value ${\bm g}^{\prime}_{rs}$, and $b_{rs}^{-1}$ is its
inverse. Finally, $J\left( {\bm v}_{r},{\bm v}_{s}\right) $ is the
Jacobian for the transformation from $\left\{ {\bm v}_{r},{\bm
v}_{s}\right\} $ to $ \left\{ {\bm v}_{r}^{\prime},{\bm
v}_{s}^{\prime}\right\}$,
\begin{equation}
J\left({\bm v}_{r},{\bm v}_{s}\right) =\left| \frac{\partial \left(
b_{rs}{\bm v}_{r},b_{rs} {\bm v}_{s}\right) }{\partial \left( {\bm
v}_{r},{\bm v}_{s}\right) }\right| ^{-1}.  \label{b.14}
\end{equation}
The delta function in (\ref{b.11}) and (\ref{b.12}) requires that
the pair is at contact, while the theta function requires that the
directions of velocities are such as to assure a collision. A
derivation of these results and further details are given in the
companion paper following this one.

\section{Homogeneous Cooling Solution}
\label{ap3}

In this Appendix, Eqs. (\ref{4.10}) and (\ref{4.11}) are proved,
leading to the exact solution of the Liouville equation
(\ref{4.12}) for homogeneous perturbations of the HCS. The HCS is
the stationary solution to the Liouville equation (\ref{4.5a}),
i.e.,
\begin{equation}
\overline{\mathcal{L}}_{T}\rho _{h}(\Gamma ;n,T,{\bm U})=0,
\label{c.1}
\end{equation}
\begin{equation}
\overline{\mathcal{L}}_{T}\equiv -\zeta _{0}\left( n,T\right) T
\frac{\partial}{\partial T}+ \overline{L}.  \label{c.2}
\end{equation}
The action of the operator $\overline{\mathcal{L}}_{T}$ on $\Psi
_{\alpha }$ can be evaluated as follows:
\begin{eqnarray}
\overline{\mathcal{L}}_{T}\Psi _{\alpha }\left( \Gamma ;n,T,{\bm
U} \right) &=& \left( \frac{\partial \left[
\overline{\mathcal{L}}_{T}\rho _{h}\left( \Gamma ;n,T,{\bm
U}\right) \right] }{\partial y_{\alpha }}\right)_{y_{\beta \neq
\alpha}}+ \left( \frac{\partial \left[ \zeta _{0} \left(
n,T\right) T\right] }{
\partial y_{\alpha }} \right)_{y_{\beta \neq \alpha}} \frac{\partial \rho _{h}
\left( \Gamma ;n,T,{\bm U}
\right)}{\partial T}  \nonumber  \\
&=& \left( \frac{\partial \left[ \zeta _{0}\left( n,T\right)
T\right] }{\partial
y_{\alpha }} \right)_{y_{\beta \neq \alpha}} \Psi _{2}\left( \Gamma ;n,T,{\bm U}\right)  \nonumber \\
&=& \sum_{\beta} \Psi _{\beta }\left( \Gamma ;n,T,{\bm U}\right)
\mathcal{K}_{\beta \alpha }^{hyd}(n,T;\mathbf{0}),  \label{c.5}
\end{eqnarray}
where $\mathcal{K}_{\beta \alpha }^{hyd}(n,T;{\bm 0})$ has been
identified in the last line from Eqs.\ (\ref{2.14})-(\ref{2.15})
particularized for ${\bm k}={\bm 0}$. This proves Eq.\
(\ref{4.10}).

Next, using this result,
\begin{eqnarray}
\overline{\mathcal{L}}^{2}\Psi_{\alpha}(\Gamma;n,T,{\bm U}) & = &
\sum_{\beta} \left[ \overline{\mathcal{L}}_{T} \Psi_{\beta}
(\Gamma;n,T,{\bm U}) \right] \mathcal{K}_{\beta \alpha}^{hyd}
(n,T;{\bm 0})  \nonumber \\
&& +\sum_{\beta} \Psi_{\beta} (\Gamma;n,T,{\bm U})
\overline{\mathcal{L}}_{T} \mathcal{K}_{\beta \alpha}^{hyd}
(n,T;{\bm 0}) \nonumber \\
& = & \sum_{\beta} \sum_{\gamma} \Psi_{\gamma}(\Gamma; n,T,{\bm
U}) \mathcal{K}_{\gamma \beta}^{hyd}(n,T;{\bm 0})
\mathcal{K}_{\beta
\alpha}^{hyd}(n,T;{\bm 0})  \nonumber \\
& & + \sum_{\beta} \Psi_{\beta} (\Gamma;n,T,{\bm U}) \left[ -
\zeta_{0}(n,T) T \frac{\partial}{\partial T}\right]
\mathcal{K}_{\beta \alpha}^{hyd}(n,T;{\bm 0}) \nonumber \\
&=& \sum_{\beta} \sum_{\gamma} \Psi_{\gamma}(\Gamma; n,T,{\bm U})
\left[ \mathcal{K}_{\gamma \beta}^{hyd}(n,T;{\bm 0})-
\delta_{\gamma \beta} \zeta_{0}(n,T) T \frac{\partial}{\partial T}
\right] \mathcal{K}_{\beta \alpha}^{hyd}(n,T;{\bm 0}),
\nonumber \\
\label{c.6}
\end{eqnarray}
or, in a compact matrix notation,
\begin{equation}
\overline{\mathcal{L}}^{2}_{T} \Psi= \Psi \left( \mathcal{K}^{hyd}-
I \zeta_{0} T \frac{\partial}{\partial T} \right) \mathcal{K}^{hyd},
\label{c.7}
\end{equation}
with $I$ denoting here the $d+2$ unit matrix. Applying the induction
method,
\begin{eqnarray}
\overline{\mathcal{L}}_{T} ^{l}\Psi &= & \overline{\mathcal{L}}_{T}
\left[ \Psi  \left( \mathcal{K}^{hyd} - I \zeta_{0} T
\frac{\partial}{\partial T} \right)^{l-2}
\mathcal{K}^{hyd} \right] \nonumber \\
& = & \left( \overline{\mathcal L}_{T} \Psi \right)\left(
\mathcal{K}^{hyd} - I \zeta_{0} T \frac{\partial}{\partial T}
\right)^{l-2} \mathcal{K}^{hyd}  + \Psi \overline{\mathcal L}_{T}
\left[\left( \mathcal{K}^{hyd} - I \zeta_{0} T
\frac{\partial}{\partial T} \right)^{l-2}
\mathcal{K}^{hyd} \right] \nonumber \\
&=& \left( \Psi \mathcal{K}^{hyd} \right) \left( \mathcal{K}^{hyd} -
I \zeta_{0} T \frac{\partial}{\partial T} \right)^{l-2}
\mathcal{K}^{hyd}  + \Psi \left( - \zeta_{0} T
\frac{\partial}{\partial T} \right) \left( \mathcal{K}^{hyd} - I
\zeta_{0} T
\frac{\partial}{\partial T} \right)^{l-2} \mathcal{K}^{hyd} \nonumber \\
&=& \Psi \left(  \mathcal{K}^{hyd} - I \zeta_{0} T
\frac{\partial}{\partial T} \right)^{l-1} \mathcal{K}^{hyd} \nonumber \\
& = & \Psi \left(  \mathcal{K}^{hyd} - I \zeta_{0} T
\frac{\partial}{\partial T} \right)^{l}. \label{c.8}
\end{eqnarray}
This implies
\begin{eqnarray}
e^{-\overline{\mathcal{L}}_{T}t}\Psi _{\alpha }\left( \Gamma
;n,T,{\bm U} \right) &=& \sum_{\beta} \Psi _{\beta }\left( \Gamma
;n,T,{\bm U} \right) \left\{ \exp - \left[
\mathcal{K}^{hyd}(n,T;{\bm 0})- I \zeta _{0}\left( n,T\right)
T \frac{\partial}{\partial T}\right] t \right\}_{\beta \alpha }  \nonumber \\
&=& \sum_{\beta} \Psi _{\beta }\left( \Gamma ;n,T,{\bm U}\right)
\widetilde{C}_{\beta \alpha }^{hyd}\left( n,T;{\bm 0},t\right),
\label{c.8a}
\end{eqnarray}
which proves Eq.\ (\ref{4.11}). In the last transformation, the
formal solution of Eq.\ (\ref{2.21}) has been used.

\section{Microscopic Conservation Laws (Balance Equations)}
\label{ap4}

\subsection{Fluxes associated with $\widetilde{a}_{\protect\alpha }(\Gamma
;n,T;{\bm k})$}

The microscopic balance equations for the phase functions
$\widetilde{a} _{\alpha }(\Gamma ;n,T;{\bm k})$ follow from those
for the Fourier transformed number density
$\widetilde{\mathcal{N}}(\Gamma;{\bm k})$, energy density
$\widetilde{\mathcal{E}}( \Gamma;{\bm k})$, and momentum density
$\widetilde{\bm{\mathcal{G}}}(\Gamma;{\bm k}) $ defined in Eq.\
(\ref {6.1}). These balance equations relate the time dependence of
the densities to appropriate fluxes
\begin{equation}
\partial _{t}e^{Lt}\left(
\begin{array}{c}
\widetilde{\mathcal{N}} \left( \Gamma ;{\bm k}\right) \\
\widetilde{\mathcal{E}} \left( \Gamma ;{\bm k}\right) \\
\widetilde{\bm{\mathcal G}} \left( \Gamma ; {\bm k} \right)
\end{array}
\right) =i {\bm k}\cdot e^{Lt}\left(
\begin{array}{c}
\frac{\widetilde{\bm{\mathcal G}}\left( \Gamma ;{\bm k} \right)}{m} \\
\widetilde{\bm s} \left( \Gamma ;{\bm k} \right) \\
\widetilde{\sf h} \left( \Gamma ; {\bm k} \right)
\end{array}
\right) -e^{Lt}\left(
\begin{array}{c}
0 \\
\widetilde{w}\left(\Gamma;{\bm k}\right) \\
0
\end{array}
\right) .  \label{d.0}
\end{equation}
These are microscopic conservation laws for
$\widetilde{\mathcal{N}}\left(\Gamma; {\bm k}\right) $ and
$\widetilde{\mathcal{\bm G}}\left(\Gamma; {\bm k} \right) $. For
granular fluids, the energy density has a source $\widetilde{w}
\left( \Gamma ;{\bm k} \right) $ due to the inelasticity of the
collisions. The forms of the fluxes of
$\widetilde{\mathcal{N}}\left(\Gamma; {\bm k} \right) $ and
$\widetilde{\bm{\mathcal G}}\left( {\bm k}\right) $ are obtained
from
\begin{equation}
L\widetilde{\mathcal{N}}\left( \Gamma ;{\bm k} \right) =\frac{i}{m}
{\bm k}\cdot \widetilde{\bm{\mathcal G}} \left( \Gamma ;{\bm k}
\right) ,\quad L\widetilde{\bm{\mathcal G}} \left( \Gamma ; {\bm k}
\right) =i{\bm k} \cdot \widetilde{\sf h}\left( \Gamma ;{\bm k}
\right) . \label{d.1}
\end{equation}
The expression for the tensor momentum flux $\widetilde{\sf h} $ is
\begin{equation}
\widetilde{\sf h}_{ij}\left( \Gamma ;{\bm k} \right)
=\sum_{r=1}^{N}m v_{r, i}v_{r, j}e^{i {\bm k} \cdot {\bm
q}_{r}}+\frac{1}{2} \int_{0}^{1}dx\, \sum_{r=1}^{N} \sum_{s \neq
r}^{N} q_{rs, i }F_{j}\left( {\bm q}_{rs},{\bm g} _{rs}\right) e^{i
{\bm k\cdot }\left( x {\bm q}_{rs }+ {\bm q} _{s}\right) }.
\label{d.2}
\end{equation}
This is the usual result for nonsingular forces ${\bm F}$,
generalized here to include a nonconservative contribution as well.
Some examples are discussed in Appendix \ref{ap2}. In all of this
Appendix only nonsingular forces are considered. The corresponding
results for hard spheres are given in the following companion paper.

The right sides of Eqs. (\ref{d.1}) are proportional to $k$, indicating that
they are densities of conserved variables. For the energy density there is
both a flux and a source
\begin{equation}
L \mathcal{E}\left( \Gamma ;{\bm k}\right) =i {\bm k}\cdot
\widetilde{\bm s}\left( \Gamma ; {\bm k}\right) -\widetilde{w}
\left( \Gamma ;{\bm k}\right) .  \label{d.3}
\end{equation}
The energy flux is given by
\begin{eqnarray}
\widetilde{\bm s} \left( \Gamma ; {\bm k}\right)
&=&\sum_{r=1}^{N}\left[ \frac{mv_{r}^{2}}{2}+\frac{1}{2}\sum_{s \neq
r}^{N}V \left( q_{rs}\right) \right] {\bm v}_{r}  e^{i{\bm k} \cdot {\bm q}_{r}}  \nonumber \\
&&+ \frac{1}{4}\int_{0}^{1}dx\, \sum_{r=1}^{N} \sum_{s \neq r}^{N}
{\bm q}_{rs}\left( {\bm v}_{r}+{\bm v}_{s}\right) \cdot {\bm
F}\left( {\bm q}_{rs},{\bm g}_{rs}\right) e^{i{\bm k} \cdot \left(
x\mathbf{q}_{rs}+{\bm q}_{s}\right)}\,   \label{d.4}
\end{eqnarray}
and the source term is
\begin{equation}
\widetilde{w}\left( \Gamma ;{\bm k}\right) =-\frac{1}{2}
\sum_{r=1}^{N} \sum_{s \neq r}^{N} {\bm g}_{rs}\cdot {\bm
F}^{nc}\left({\bm q}_{rs}, {\bm g}_{rs} \right) e^{i {\bm k} \cdot
{\bm q}_{r}},  \label{d.5}
\end{equation}
where ${\bm F}^{nc} ({\bm q}_{rs},{\bm g}_{rs})$ is the
nonconservative part of the force. The functional forms for the
fluxes $\widetilde{\bm G}\left( \Gamma ;{\bm k}\right) ,$
$\widetilde{\bm s}\left( \Gamma ; {\bm k}\right)$ , and
$\widetilde{\sf h} \left( \Gamma ; {\bm  k} \right) $ are the same
as those for a normal fluid, except that the total force, including
its nonconservative part, occurs. The source $\widetilde{w}\left(
\Gamma ;{\bm k}\right) $ depends only on the nonconservative part of
the force. For the special case of the force given in Eq.\
(\ref{b.1}), Eq.\ (\ref{d.5}) becomes
\begin{equation}
\widetilde{w}\left( \Gamma ;{\bm k}\right) =\frac{1}{2}
\sum_{r=1}^{N} \sum_{s \neq r}^{N} \Theta \left( \sigma
-q_{rs}\right) \gamma \left( \sigma -q_{rs}\right) \left( {\bm
g}_{rs} \cdot \widehat{\bm q}_{rs}\right) ^{2}e^{i {\bm k} \cdot
{\bm q}_{r}}, \label{d.6}
\end{equation}
which agrees with Eq.\ (\ref{b.5}) for ${\bm k}=0$.

The corresponding fluxes associated to the $\widetilde{a}_{\alpha
}(\Gamma ;n,T; {\bm k})$ follow from their definition, Eq.\
(\ref{6.03}), in terms of the above densities,
\begin{equation}
L\widetilde{a}_{\alpha }(\Gamma ;n,T;{\bm k})=i{\bm k} \cdot
\widetilde{\bm f}_{\alpha }(\Gamma ;n,T;{\bm k})-\frac{1}{e_{0,T}}
\widetilde{w} (\Gamma;{\bm k} ) \delta_{\alpha 2}, \label{d.7}
\end{equation}
with
\begin{equation}
\widetilde{\bm f}_{1}(\Gamma ;n,T;{\bm
k})=\frac{\widetilde{\bm{\mathcal G}} (\Gamma ;{\bm k})}{m},
\label{d.8}
\end{equation}
\begin{equation}
\widetilde{\bm f}_{2}(\Gamma ;n,T;{\bm k})=\frac{1}{e_{0,T}}\left[
\widetilde{\bm s} \left( \Gamma ;{\bm k} \right) -
\frac{e_{0,n}}{m}\widetilde{\bm{\mathcal G}} \left( \Gamma ;{\bm
k}\right) \right], \label{d.10}
\end{equation}
\begin{equation}
\widetilde{\sf f}_{3,ij}(\Gamma ;n,T;{\bm k})= \frac{\widetilde{\sf
h}_{ij}\left( \Gamma; {\bm k} \right)}{n m}. \label{d.8a}
\end{equation}
The last equation above gives the tensor flux associated to the
vector ${\bm a}_{3}=\bm{\mathcal{G}}/nm$.

Next, calculate the quantity $\widetilde{\ell}(\Gamma;n,T;{\bm
k})$ appearing in  Eq. (\ref{5.3}). Use of Eq.\ (\ref{5.2}) gives
directly Eq. (\ref{5.3}) with
\begin{eqnarray}
\widetilde{\ell}\left( \Gamma ;n,T; {\bm k}\right) &=&\frac{1}{
e_{0,T}}\widetilde{w}\left( \Gamma ;{\bm k}\right) +\zeta
_{0}\left( n,T\right) T \frac{\partial}{\partial T}
\widetilde{a}_{2}(\Gamma ;n,T;{\bm k} )- \sum_{\beta}
\mathcal{K}_{2\beta }^{hyd}\left( n,T;{\bm 0}\right) \widetilde{a}
_{\beta }(\Gamma ;n,T;{\bm k}) \nonumber \\
&=& \frac{1}{e_{0,T}}\widetilde{w}\left( \Gamma ;{\bm k}\right) -
\frac{\zeta _{0}(n,T) T}{e_{0,T}}\, \frac{\partial
e_{0,T}}{\partial T} \widetilde{a}_{2}(\Gamma ;n,T;{\bm k})-
\frac{\zeta _{0}(n,T) T}{e_{0,T}}\, \frac{
\partial e_{0,n}}{\partial T}\widetilde{a}_{1}(\Gamma ;n,T;
{\bm k})  \nonumber \\
&&-\frac{\partial \left[ \zeta _{0}(n,T)T \right]}{\partial
n}\widetilde{a} _{1}(\Gamma ;n,T;{\bm k})-\frac{\partial \left[
\zeta _{0}(n,T) T\right] }{
\partial T}\widetilde{a}_{2}(\Gamma ;n,T;{\bm k}) \nonumber \\
&=&\frac{1}{e_{0,T}}\left\{  \widetilde{w}\left( \Gamma ;{\bm k}
\right) -\left( \frac{\partial \left[ e_{0,T}\zeta _{0}(n,T)T
\right]}{\partial n}\right)_{T} \widetilde{a}_{1}(\Gamma ;n,T;{\bm
k})  \right. \nonumber \\
&& \left. -\left( \frac{\partial \left[e_{0,T}\zeta
_{0}(n,T)T\right]}{\partial T}\right)_{n} \widetilde{a}_{2}(\Gamma
;n,T;{\bm k})\right\}. \label{d.12}
\end{eqnarray}
From the expression for the cooling rate in the HCS given in Eq.\
(\ref{4.3d}),
\begin{equation}
e_{0,T}\zeta _{0}(n,T)T=-V^{-1}\int d\Gamma \rho _{h}(\Gamma
;n,T)LE(\Gamma) = V^{-1}\int d\Gamma \rho _{h}(\Gamma
;n,T)\widetilde{w}\left( \Gamma ; {\bm 0}\right)  \label{d.13}
\end{equation}
and, since $\widetilde{w}\left( \Gamma ;{\bm 0}\right) $ is
independent of $n$ and $T$ as seen from its definition in Eq.\,
(\ref{d.0}),
\begin{eqnarray}
\left( \frac{\partial\left( e_{0,T}\zeta _{0}T \right)}{\partial
n}\right)_{T} & = & V^{-1}\int d\Gamma\,  \Psi _{1}(\Gamma
;n,T)\widetilde{w}\left( \Gamma ;{\bm 0}\right),
\nonumber \\
\left( \frac{\partial \left( e_{0,T}\zeta _{0}T \right)}{\partial T}
\right)_{n}& = & V^{-1} \int d\Gamma\,  \Psi _{2}(\Gamma
;n,T)\widetilde{w}\left( \Gamma ;{\bm 0}\right).  \label{d.14}
\end{eqnarray}
Substitution of the above relations into Eq.\ (\ref{d.12}), noting
that the sum can be extended to include $\alpha=3$, since the new
contribution vanishes by symmetry as a consequence of
$\widetilde{w}$ being a scalar, gives Eq.\ (\ref{5.4}) in the text.

\subsection{Fluxes associated with $\widetilde{\protect\psi }_{\protect\alpha
}(\Gamma ;n,T;{\bm k})$}

By definition in Eq.\ (\ref{6.01}), the set of $\psi _{\alpha
}(\Gamma ;n,T; {\bm r})$ are densities associated with the
invariants, i.e.
\begin{equation}
\widetilde{\psi }_{\alpha }(\Gamma ;n,T;{\bm 0})=\Psi _{\alpha
}\left( \Gamma ;n,T \right),  \label{d.16}
\end{equation}
and so
\begin{equation}
\sum_{\beta} \mathcal{U} _{\alpha \beta }\left( t,T\right)
\widetilde{\psi }_{\beta }(\Gamma ;n,T; {\bm 0}) = \widetilde{\psi
}_{\alpha }(\Gamma ;n,T; {\bm 0})  \label{d.17}
\end{equation}
and
\begin{equation}
\partial _{t}\widetilde{\psi }_{\alpha }(\Gamma ;n,T;{\bm 0},t) =
0.
\label{d.17aa}
\end{equation}

 The time derivative of $\widetilde{\psi }_{\alpha }(\Gamma
;n,T;{\bm k} ,t) $ must be of order $k$, so there exists a flux
$\widetilde{\bm \gamma }_{\beta }\left( \Gamma ;n,T;{\bm k},t\right)
$ such that
\begin{equation}
\partial _{t}\widetilde{\psi }_{\alpha }(\Gamma ;n,T;{\bm
k},t)-i{\bm k\cdot }\widetilde{\bm \gamma }_{\alpha }\left( \Gamma
;n,T;{\bm k} ,t\right) =0.  \label{d.18}
\end{equation}
The generator of the dynamics $\mathcal{U}\left( t,T\right) $ is
defined by Eq.\ (\ref{4.14}), and taking the time derivative there,
it is seen to obey the equation
\begin{equation}
\partial _{t}\mathcal{U} \left( t,T \right) =-\left[ \overline{\mathcal{L}}_{T}
-\mathcal{K}^{hyd \text{ T}}(n,T;{\bm 0})\right] \mathcal{U}
\left( t,T\right) , \label{d.20}
\end{equation}
where $\mathcal{K}^{hyd \text{ T}}$ is the transpose of
$\mathcal{K}^{hyd} $. This equation can be formally integrated to
write
\begin{equation}
\mathcal{U}(t,T)= \exp \left\{ - t \left[ \overline{\mathcal{L}}_{T}
-\mathcal{K}^{hyd \text{ T}}(n,T;{\bm 0}) \right] \right\},
\label{d.21a}
\end{equation}
that shows that Eq.\ (\ref{d.20}) is equivalent to
\begin{equation}
\partial _{t}\mathcal{U} \left( t,T \right) =-\mathcal{U}
\left( t,T\right) \left[ \overline{\mathcal{L}}_{T}
-\mathcal{K}^{hyd \text{ T}}(n,T;{\bm 0})\right]  . \label{d.21b}
\end{equation}
Use of this into Eq.\ (\ref{5.5aaa}) yields
\begin{equation}
\partial_{t} \widetilde{\psi }(\Gamma ;n,T;{\bm k},t)=-
\mathcal{U}(t,T) \left[\overline{\mathcal{L}}_{T}
-\mathcal{K}^{hyd \text{ T}}(n,T;{\bm 0})\right]\widetilde{\psi
}(\Gamma ;n,T;{\bm k}). \label{d.22}
\end{equation}
Comparison with Eq.\ (\ref{d.18}) leads to the identifications given
in Eqs. (\ref{5.5ab}) and (\ref{5.5ac}).

\section{Details of Euler Order Parameters}
\label{ap5} The Euler order time independent correlation function
$\widehat{\bm k} \cdot \overline{\bm D}_{\alpha \beta }(n,T;{\bm
0},0)$ whose expression is given in Eq.\  (\ref{5.12a}), is
determined from direct evaluation. Consider first the case $\alpha
=1$ for which $ {\bm f}_{1}(\Gamma ;n,T;{\bm
0})=\widetilde{\bm{\mathcal{G}}} \left( \Gamma ; {\bm 0}\right)/m $.
Then
\begin{eqnarray} \widehat{\bm k} \cdot
\overline{\bm D}_{1\beta }(n,T;{\bm 0},0) &=&(mV) ^{-1} \int d\Gamma
\widehat{\bm k} \cdot \widetilde{\bm{\mathcal{G}}}(\Gamma;{\bm 0})
\left[ \frac{\partial \rho_{h}(\Gamma;n,T,{\bm U})}{\partial
y_{\beta}} \right]_{{\bm
U}={\bm 0}} \nonumber \\
&=& (mV)^{-1} \left[ \frac{\partial}{\partial y_{\beta}} \int d
\Gamma\, \widehat{\bm k} \cdot \widetilde{\bm{\mathcal{G}}}
(\Gamma;{\bm 0})
\rho_{h}(\Gamma;n,T,{\bm U}) \right]_{{\bm U}={\bm 0}} \nonumber \\
& = & V^{-1} \left[ \frac{\partial}{\partial y_{\beta}} N
\widehat{\bm k} \cdot {\bm U} \right]_{{\bm U}={\bm 0}} \nonumber
\\
&=& \delta_{\beta 3} n, \label{e.1}
\end{eqnarray}
in agreement with Eq.\ (\ref{5.12}).  For $\alpha =2$, use Eq.\
(\ref{d.10}) to get
\begin{eqnarray}
\widehat{\bm k}\cdot \overline{\bm D}_{2\beta }(n,T;{\bm 0},0) & = &
(V e_{0,T})^{-1} \int d\Gamma\,  \widehat{\bm k} \cdot \left[
\widetilde{\bm s}(\Gamma;{\bm 0}) -\frac{e_{0,n}}{m}
\widetilde{\bm{\mathcal{G}}} (\Gamma;0) \right] \left[
\frac{\partial \rho_{h}
(\Gamma;n,t,{\bm U})}{\partial y_{\beta}} \right]_{{\bm U}={\bm 0}} \nonumber \\
& = & (V e_{0,T})^{-1} \left[ \frac{\partial}{\partial y_{\beta}}
\int d\Gamma\, \widehat{\bm k} \cdot \widetilde{\bm s}(\Gamma;{\bm
0}) \rho_{h}(\Gamma;n,T,{\bm U}) \right]_{{\bm U}={\bm 0}}
\nonumber \\
&&  - e_{0,n} (mVe_{0,T})^{-1} \left[ \frac{\partial}{\partial
y_{\beta}} \int d\Gamma\, \widehat{\bm k} \cdot
\widetilde{\bm{\mathcal G}}(\Gamma;{\bm 0})
\rho_{h}(\Gamma;n,T,{\bm U}) \right]_{{\bm U}={\bm 0}}.
\label{e.2}
\end{eqnarray}
The second term on the right hand side is easily evaluated using the
result in Eq.\ (\ref{e.1}). The ensemble average in the first term
can be carried out by making the change of velocity variables ${\bm
v}_{r} \rightarrow {\bm v}_{r}+{\bm U}$ and using the properties of
the Galilean transformation,
\begin{eqnarray}
\int d \Gamma\, \widetilde{\bm s} (\Gamma;{\bm 0})
\rho_{h}(\Gamma;n,T,{\bm U}) & = & \int d \Gamma \left[
\widetilde{\bm s} (\Gamma;{\bm 0})\right]_{\{ {\bm v}_{r}
\rightarrow {\bm v}_{r}+{\bm U} \}} \rho_{h}(\Gamma;n,T,{\bm
U}={\bm 0})
\nonumber \\
& = & \int d\Gamma\, \left\{ \widetilde{\bm s} (\Gamma;{\bm 0})
+\left[ \mathcal{E}(\Gamma;{\bm 0})+ \frac{1}{2} m N U^{2} \right]
{\bm U} + \widetilde{\sf h} (\Gamma;{\bm 0}) \cdot {\bm U} \right.
\nonumber \\
& & + \left. \frac{1}{2} \widetilde{\bm{\mathcal{G}}} (\Gamma;{\bm
0}) U^{2} + {\bm U}{\bm U} \cdot \widetilde{\bm{\mathcal{G}}}
(\Gamma;{\bm 0})
\right\} \rho_{h}(\Gamma;n,T) \nonumber \\
&=& e_{0}(n,T) V {\bm U}+ \frac{1}{2} m N U^{2} {\bm U} + \int
d\Gamma \widetilde{\sf h} (\Gamma;{\bm 0}) \cdot {\bm U}
\rho_{h}(\Gamma;n,T).  \nonumber \\ \label{e.3}
\end{eqnarray}
Using this result it is easily obtained
\begin{equation}
\widehat{\bm k} \cdot \overline{\bm D}_{2\beta }(n,T;{\bm 0},0)=
\frac{\delta_{\beta 3}}{e_{0,T}} \left[ e_{0}-e_{0,n} n + (Vd)^{-1}
\int d\Gamma  \text{tr } {\sf H} (\Gamma) \rho_{h} (\Gamma;n,T)
\right]. \label{e.4}
\end{equation}
Here ${\sf H}(\Gamma)  \equiv {\sf h}(\Gamma,{\bm 0})$, so $\text{tr
} {\sf H} (\Gamma) = \sum_{i=1}^{d} \widetilde{\sf
h}_{ii}(\Gamma;{\bm 0})$ is the volume integrated momentum flux. For
the case of the dissipative hard spheres discussed in Appendix
\ref{ap2}, it follows from Eq.\ (\ref{d.2}) that
\begin{equation}
 {\sf H}_{ij}\left( \Gamma \right) =\sum_{r=1}^{N}mv_{r,i} v_{r,j}+\frac{1}{2}
\sum_{r}^{N} \sum_{s \neq r}^{N} {q}_{rs,i}F_{j}\left( {\bm
q}_{rs},{\bm g}_{rs}\right) . \label{e.6a}
\end{equation}

For $\alpha =3$, use Eq.\ (\ref{d.8a}) to get
\begin{eqnarray}
\widehat{\bm k} \cdot \overline{\bm D}_{3\beta }(n,T;{\bm 0},0)
&=&\left( nmV\right)^{-1} \left[ \frac{\partial }{\partial y_{\beta
}}\int d\Gamma\, {\sf h}_{\parallel \parallel}(\Gamma;{\bm 0})\rho
_{h}\left( \Gamma ;n,T,{\bm U}\right) \right]_{{\bm U}={\bm 0}}
\nonumber \\
&=& \left( nmV\right)^{-1} \left\{ \frac{\partial }{\partial
y_{\beta }} \int d\Gamma\,  \left[ {\sf h}_{\parallel \parallel}
(\Gamma;{\bm 0})+ 2 U_{\parallel} \mathcal{G}_{\parallel}
(\Gamma;{\bm 0}) \right. \right. \nonumber
\\
& & \left. \left. + m n U_{\parallel}^{2} \right] \rho _{h}\left(
\Gamma ;n,T,{\bm U}={\bm 0}\right) \right\}_{{\bm U}={\bm 0}},
\label{e.5}
\end{eqnarray}
where again the change of velocity variables has been made. The
subindex $\parallel$ indicates the component in the direction of
$\widehat{\bm k}$. Therefore,
\begin{equation}
\widehat{\bm k} \cdot \overline{\bm D}_{3\beta }(n,T;{\bm
0},0)=(mn)^{-1}  \left( \delta _{\beta 1} \frac{\partial}{\partial
n}+\delta _{\beta 2} \frac{\partial}{\partial T} \right) (Vd)^{-1}
\int d\Gamma\, \text{tr } {\sf H} (\Gamma) \rho _{h}\left( \Gamma
;n,T\right). \label{e.6}
\end{equation}

These results are consistent with the form of the phenomenological
matrix $\mathcal{K}^{hyd,(a)}_{1}$ in Eq.\ (\ref{5.12r}), if the
pressure is identified as
\begin{equation}
p(n,T)\equiv  (Vd)^{-1} \int d\Gamma\,  \text{tr} {\sf H} \left(
\Gamma \right) \rho _{h}\left( \Gamma ;n,T\right). \label{e.7}
\end{equation}
This is Eq.\ (\ref{5.12b}) in  the main text.

The single transport coefficient at Euler order $\zeta ^{U}\left(
n,T\right) $ was identified in Eq.\ (\ref{5.8b}) as
\begin{equation}
T\zeta ^{U}(n,T)=- \lim \widehat{\bm k} \cdot \overline{\bm
S}_{23}^{(1)}(n,T;t), \label{e.8}
\end{equation}
where the expression for ${\bm S}_{23}^{(1)}$ follows from Eq.\
(\ref{5.5f}),
\begin{eqnarray}
\overline{\bm S}_{23}^{(1)}(n,T;t) &=&V^{-1}\int d\Gamma\,
\widetilde{\bm \ell }^{(1)}(\Gamma ;n,T) \Psi _{3}(\Gamma
;n,T)  \nonumber \\
&&-  V^{-1} \int d\Gamma\, \widetilde{\ell}(\Gamma ;n,T;{\bm 0})
\widetilde{\bm{\psi }}_{3 }^{(1)}(\Gamma ;n,T;t) . \label{e.9}
\end{eqnarray}
The first term on the right hand side  vanishes since
\begin{equation}
\int d\Gamma \widetilde{\bm{\ell }}^{(1)}(\Gamma ;n,T)\Psi
_{3}(\Gamma ;n,T)= \left[ \frac{\partial}{\partial U_{\parallel}}
\int d\Gamma\, \widetilde{\bm{\ell}}^{(1)}(\Gamma;n,T) \rho
_{h}\left( \Gamma ;n,T,{\bm U}\right) \right]_{{\bm U}={\bm 0}} .
\label{e.10}
\end{equation}
The same change of variables ${\bm v}_{r}\rightarrow {\bm v} _{r} +
{\bm U}$ as above, shows that this vanishes due to spherical
symmetry of $\rho _{h}\left( \Gamma ;n,T,{\bm U}={\bm 0}\right) $.
The remaining contribution to $\zeta^{U}(n,T)$ as given by Eq.\
(\ref{e.8}), is made more explicit using the definition of
$\widetilde{\bm{\ell} }(\Gamma ;n,T;{\bm 0})$, Eq. (\ref{5.4b}), and
also that of $\mathcal{U}_{3\alpha }\left( t,T\right)$, Eq.
(\ref{4.14}), to compute
\begin{equation}
\widetilde{\bm \psi}_{3}^{(1)}(\Gamma;n,T;t)=\sum_{\alpha}
\mathcal{U}_{3 \alpha}(t,T) \widetilde{\bm
\psi}_{3}^{(1)}(\Gamma;n,T). \label{e.10a}
\end{equation}
This gives
\begin{eqnarray}
\zeta ^{U}(n,T) &= &  +\lim (V T )^{-1} \sum_{\alpha} \int d\Gamma\,
\widetilde{l}(\Gamma;n,T;{\bm 0}) \mathcal{U}_{3 \alpha} (t,T)
\widehat{\bm k} \cdot
\widetilde{\bm{\psi}}_{\alpha}^{(1)}(\Gamma;n,T)
\nonumber \\
& = & - \lim (VT e_{0,T})^{-1} \sum_{\alpha} \int d\Gamma\, \left[
\left( 1-P^{\dagger }\right) LE(\Gamma) \right] \delta_{\alpha 3}
e^{-\overline{\mathcal{L}}_{T}t} \widehat{\bm k} \cdot
\widetilde{\bm{\psi}}^{(1)}_{\alpha}(\Gamma;n,T)
\nonumber \\
& =&\lim (VTe_{0,T}d)^{-1} \int d\Gamma\,  W\left( \Gamma
;n,T\right) e^{-\overline{\mathcal{L}}_{T}t}\mathcal{M}_{\zeta
^{U}}\left( \Gamma ;n,T\right). \label{e.11}
\end{eqnarray}
The phase function $W\left( \Gamma ;n,T\right) $ is defined by
\begin{eqnarray}
W\left( \Gamma ;n,T\right) &=&- \left( 1-P^{\dagger }\right) LE (\Gamma)  \nonumber \\
&=& - LE(\Gamma)  - N \left( \frac{\partial}{\partial n} \left(
e_{0,T}T\zeta _{0}\right) \right)_{e_{0}}  - E(\Gamma) \left(
\frac{\partial}{\partial e_{0}} \left( e_{0,T}T\zeta _{0}\right)
\right)_{n}.  \label{e.12}
\end{eqnarray}
The second equality  follows from explicitly evaluating the action
of the projection operator $P^{\dagger }$, defined in Eq.\
(\ref{5.4a}), on $LE(\Gamma)$. Finally, the phase function $
\mathcal{M}_{\zeta ^{U}}\left( \Gamma ;n,T\right) $ is
\begin{eqnarray}
\mathcal{M}_{\zeta ^{U}}\left( \Gamma ;n,T\right) & =  & d
\widehat{\bm k} \cdot
\widetilde{\bm{\psi}}^{(1)}_{3}(\Gamma;n,T)  d \nonumber \\
& = & d \int d{\bm r} r_{\parallel} \left[ \frac{\delta
\rho _{\ell h}}{\delta U_{ \parallel}\left({\bm r}\right) } \right]_{\{y_{\beta}\}=\{n,T,{\bm 0}\}}
\nonumber \\
& = & -\sum_{r=1}^{N}{\bm q}_{r} \cdot \frac{\partial
\rho_{h}(\Gamma;n,T)}{\partial {\bm v}_{r}}. \label{e.13}
\end{eqnarray}
In the last transformation, the local equilibrium form for the
velocity dependence has been taken into account,
\begin{equation}
\left[ \rho _{lh}\left( \Gamma | \left\{ y_{\beta }\right\} \right)
\right]_{ \{y_{\beta}\} =\{n,T,{\bm U}({\bm r}) \}} =\rho _{h}\left[
\left\{ {\bm q}_{r},{\bm v}_{r}- {\bm U}({\bm q}_{r})\right\};n,T
\right].  \label{e.14}
\end{equation}

\section{Navier-Stokes Order Transport Coefficients}
\label{ap6}  The Helfand forms for the transport coefficients are
identified from (\ref {5.8c}) and (\ref{5.8d}). For the energy flux,
these are the thermal conductivity $\lambda $ and the new granular
fluid coefficient $\mu $,
\begin{equation}
\lambda = e_{0,T}\lim \widehat{\bm k} \widehat{\bm k} : \left[
\overline{\sf D} _{22}^{(1)}(n,T;t)- \sum_{\alpha} \overline{\bm
D}_{2 \alpha}(n,T;{\bm 0}, 0)\overline{\bm C}_{\alpha
2}^{(1)}(n,T;t)\right] , \label{f.1}
\end{equation}
\begin{equation}
\mu = e_{0,T}\lim \widehat{\bm k} \widehat{\bm k} : \left[
\overline{\sf D} _{21}^{(1)}(n,T;t)- \sum_{\alpha} \overline{\bm
D}_{2 \alpha}(n,T;{\bm 0}, 0)\overline{\bm C}_{\alpha
1}^{(1)}(n,T;t)\right]   ,  \label{f.2}
\end{equation}
The shear and bulk viscosities, $\eta $ and $\kappa $, are
identified as
\begin{equation}
\eta = m n \lim \widehat{\bm k} \widehat{\bm k} : \left[
\overline{\sf D} _{44}^{(1)}(n,T;t)- \sum_{\alpha} \overline{\bm
D}_{4 \alpha}(n,T;{\bm 0}, 0)\overline{\bm C}_{\alpha
4}^{(1)}(n,T;t)\right], \label{f.3}
\end{equation}
\begin{equation}
\kappa +\frac{2(d-1)\eta}{d} =  m n \lim \widehat{\bm k}
\widehat{\bm k} : \left[ \overline{\sf D} _{33}^{(1)}(n,T;t)-
\sum_{\alpha} \overline{\bm D}_{3 \alpha}(n,T;{\bm 0},
0)\overline{\bm C}_{\alpha 3}^{(1)}(n,T;t)\right] . \label{f.4}
\end{equation}
Finally, the two Navier-Stokes transport coefficients associated with the
cooling rate are
\begin{equation}
\zeta ^{n}= T^{-1}\lim \left[ \widehat{\bm k} \widehat{\bm k}: {\sf
S}^{(2)}_{21} (n,T;t)+T \zeta^{U}(n,T) \widehat{{\bm k}} \cdot {\bm
C}_{31}^{(1)}(n,T;t) \right], \label{f.5}
\end{equation}
\begin{equation}
\zeta ^{T}= T^{-1}\lim \left[ \widehat{\bm k} \widehat{\bm k}: {\sf
S}^{(2)}_{22} (n,T;t)+T \zeta^{U}(n,T) \widehat{{\bm k}} \cdot {\bm
C}_{32}^{(1)}(n,T;t) \right]         . \label{f.6}
\end{equation}

The corresponding Green-Kubo forms are
\begin{equation}
\lambda =e_{0,T} \widehat{\bm k} \widehat{\bm k} : \left[ {\sf
D}_{22}^{(1)}(n,T;{\bm 0})-\lim \int_{0}^{t} dt^{\prime}\, {\sf
G}_{22}(n,T;t^{\prime}) \right], \label{f.7}
\end{equation}
\begin{equation}
\mu = e_{0,T} \widehat{\bm k} \widehat{\bm k} : \left[ {\sf
D}_{21}^{(1)}(n,T;{\bm 0})-\lim \int_{0}^{t} dt^{\prime}\, {\sf
G}_{21}(n,T;t^{\prime}) \right], \label{f.8}
\end{equation}
\begin{equation}
\eta =nm  \widehat{\bm k} \widehat{\bm k} : \left[ {\sf
D}_{44}^{(1)}(n,T;{\bm 0})-\lim \int_{0}^{t} dt^{\prime}\, {\sf
G}_{44}(n,T;t^{\prime}) \right], \label{f.9}
\end{equation}
\begin{equation}
\kappa +\frac{2(d-1)\eta}{d}\ = nm  \widehat{\bm k} \widehat{\bm k}
: \left[ {\sf D}_{33}^{(1)}(n,T;{\bm 0})-\lim \int_{0}^{t}
dt^{\prime}\, {\sf G}_{33}(n,T;t^{\prime}) \right], \label{f.10}
\end{equation}
\begin{eqnarray}
\zeta ^{n} &=& T^{-1} \widehat{\bm k} \widehat{\bm k}: \left\{ {\sf
S}^{(2)}_{21}(n,T;{\bm 0})  \right. \nonumber \\
&& \left. -\lim \int_{0}^{t} dt^{\prime}  \left[ {\sf
N}_{21}^{(1)}(n,T;t^{\prime})+T \zeta^{U} (n,T) \widehat{\bm k} {\bm
E}_{31} (n,T;{\bm 0},t^{\prime}) \right] \right\}, \label{f.11}
\end{eqnarray}
\begin{eqnarray}
\zeta ^{T}  &=& T^{-1} \widehat{\bm k} \widehat{\bm k}: \left\{ {\sf
S}^{(2)}_{22}(n,T;{\bm 0})  \right. \nonumber \\
&& \left. -\lim \int_{0}^{t} dt^{\prime}  \left[ {\sf
N}_{22}^{(1)}(n,T;t^{\prime})+T \zeta^{U} (n,T) \widehat{\bm k} {\bm
E}_{32} (n,T;{\bm 0},t^{\prime}) \right] \right\}. \label{f.12}
\end{eqnarray}

\end{document}